\documentclass[12pt,letterpaper]{JHEP3}

\usepackage{amscd,amsmath,amssymb,amsfonts,xspace,mathrsfs}
\usepackage[multiple]{footmisc}

\numberwithin{equation}{section}

\def\varpi{t}

\def\det{\,{\rm det}\, }

\def\Re{\,{\rm Re}\,}

\def\({\left(}
\def\){\right)}
\def\[{\left[}
\def\]{\right]}
\def\<{\left\langle}
\def\>{\right\rangle}
\def\hf{{1\over 2}}
\def\haf{\textstyle{1\over 2}}

\newcommand{\de}{\mathrm{d}}

\newcommand{\I}{\mathrm{i}}

\newcommand{\cL}{\mathcal{L}}
\newcommand{\cD}{\mathcal{D}}

\newcommand{\p}{\partial}

\newcommand{\half}{\frac{1}{2}}

\newcommand{\cF}{\mathcal{F}}

\newcommand{\cC}{\mathcal{C}}
\newcommand{\cS}{\mathcal{S}}

\newcommand{\cK}{\mathcal{K}}
\newcommand{\cM}{\mathcal{M}}

\newcommand{\cN}{\mathcal{N}}
\newcommand{\cE}{\mathcal{E}}
\newcommand{\cX}{\mathcal{X}}

\newcommand{\cP}{\mathcal{P}}

\newcommand{\cT}{\mathcal{T}}

\newcommand{\ub}{\bar{u}}

\DeclareSymbolFont{AMSa}{U}{msa}{m}{n}
\DeclareSymbolFont{AMSb}{U}{msb}{m}{n}
\DeclareMathSymbol{\fieldR}{\mathalpha}{AMSb}{"52}

\newcommand{\kahler}{{K\"ahler}\xspace}

\newcommand{\qk}{{quaternion-K\"ahler}\xspace}

\newcommand{\cZ}{\mathcal{Z}}
\newcommand{\cI}{\mathcal{I}}

\newcommand{\cU}{\mathcal{U}}

\newcommand{\abs}[1]{\lvert#1\rvert}

\newcommand{\pa}{\partial}
\newcommand{\nn}{\nonumber}

\newcommand{\eps}{\epsilon}

\newcommand{\IR}{\mathbb{R}}
\newcommand{\IC}{\mathbb{C}}
\newcommand{\IZ}{\mathbb{Z}}

\newcommand{\sgn}{\mbox{sgn}}
\newcommand{\tzeta}{\tilde\zeta}

\newcommand{\tc}{\tilde c}

\newcommand{\txi}{\tilde\xi}

\newcommand{\CP}{\IC P^1}

\def\bea{\begin{eqnarray}}
\def\eea{\end{eqnarray}}
\def\be{\begin{equation}}
\def\ee{\end{equation}}
\def\ba{\begin{align}}
\def\ea{\end{align}}
\def\bse{\begin{subequations}}
\def\ese{\end{subequations}}

\def\ba{\bar a}

\def\bz{\bar z}
\def\btau{\bar \tau}

\def\bZ{\bar Z}

\def\bF{\bar F}

\newcommand{\CZ}{{\cal{Z}}}

\newcommand{\cB}{\mathcal{B}}

\def\cij#1{c}
\def\ci#1{c}

\def\txii#1{{\tilde\xi}^{[#1]}}

\def\ai#1{{\alpha}^{[#1]}}

\def\xii#1{\xi_{[#1]}}

\def\Hij#1{H^{[#1]}}

\def\Hij#1{H^{[#1]}}

\def\XXint#1#2#3{{\setbox0=\hbox{$#1{#2#3}{\int}$}
\vcenter{\hbox{$#2#3$}}\kern-.5\wd0}}

\newcommand{\cY}{\mathcal{Y}}

\def\Fcl{F^{\rm cl}}

\newcommand{\expe}[1]{{\bf E}\!\left( #1\right)}

\def\gamD#1{\tilde\gamma}

\def\CY{\mathfrak{Y}}
\def\CYm{{{\mathfrak{Y}}}}

\DeclareMathOperator{\Td}{Td}
\DeclareMathOperator{\ch}{ch}

\DeclareMathOperator{\Erf}{Erf}
\DeclareMathOperator{\Erfc}{Erfc}

\def\cla{\tilde c_a}
\def\cl0{\tilde c_0}

\newcommand{\bfb}{{\boldsymbol b}}
\newcommand{\bfc}{{\boldsymbol c}}

\newcommand{\bfk}{{\boldsymbol k}}

\newcommand{\bfp}{{\boldsymbol p}}
\newcommand{\bfq}{{\boldsymbol q}}

\newcommand{\bft}{{\boldsymbol t}}
\newcommand{\bfv}{{\boldsymbol v}}
\newcommand{\bfz}{{\boldsymbol z}}
\newcommand{\bfx}{{\boldsymbol x}}
\newcommand{\bfy}{{\boldsymbol y}}
\newcommand{\bfeps}{{\boldsymbol \epsilon}}
\newcommand{\bfmu}{{\boldsymbol \mu}}
\newcommand{\bfnu}{{\boldsymbol \nu}}
\newcommand{\bfrho}{{\boldsymbol \rho}}

\newcommand{\gammap}{\gamma}
\newcommand{\gammam}{-\gamma}

\def\trans{g}

\def\kbp{(k+b)_+}

\def\qbp{(q+b)_+}

\def\Erfc{\text{Erfc}}

\def\signkp{(-1)^{\bfk\cdot\bfp}}

\def\wh{\mathfrak{h}}
\def\bwh{\bar\wh}

\def\cXt{\cX^{(\theta)}}
\def\cXsf{\cX^{\rm sf}}
\def\cXcl{\cX^{\rm cl}}
\def\cXq{\cX^{(1)}}

\def\tcF{\tilde\cF}

\def\cFi#1{\cF^{(#1)}}
\def\cFq{\cF^{(1)}}

\def\hcZ{\widehat\cZ}

\def\whh{\widehat h}

\def\dInt{\cY_{\gamma_1\gamma_2}}

\def\Dela{\Delta^{\bft}}

\def\Sg{{\rm S}_{\bfp_1,\bfp_2}}
\def\Sgg{\Sg(\bfmu_1,\bfmu_2,\bfrho)}

\def\OmMSW{\Omega^{\rm MSW}}
\def\bOmMSW{{\bar \Omega}^{\rm MSW}}

\title{Multiple D3-instantons and mock modular forms I}

\preprint{L2C:16-056\\
IPhT-T16/037 \\
TCDMATH 16-08\\
CERN-TH-2016-121\\
arXiv:1605.05945v2}

\author{Sergei Alexandrov$^{1}$, Sibasish Banerjee$^2$, Jan Manschot$^3$, Boris Pioline$^{4,5}$
\\
$^1$ {\it
Laboratoire Charles Coulomb (L2C), UMR 5221 CNRS-Universit\'e de
Montpellier, F-34095, Montpellier, France}\\

$^2$ {\it IPhT, CEA, Saclay, Gif-sur-Yvette, F-91191, France}\\

$^3$ {\it School of Mathematics, Trinity College, Dublin 2, Ireland}\\

$^4$ {\it CERN PH-TH,
Case C01600, CERN, CH-1211 Geneva 23, Switzerland}\\

$^5$ {\it Laboratoire de Physique Th\'eorique et Hautes
Energies, CNRS UMR 7589, \\
Universit\'e Pierre et Marie Curie,
4 place Jussieu, 75252 Paris cedex 05, France} \\

\vspace*{2mm} {\tt e-mail:
\email{salexand@univ-montp2.fr},
\email{sibasish.banerjee@cea.fr},
\email{manschot@maths.tcd.ie},
\email{boris.pioline@cern.ch}
}

\vspace*{-3mm}

}

\abstract{
We study D3-instanton corrections to the hypermultiplet moduli space in type IIB string theory compactified
on a Calabi-Yau threefold. In a previous work, consistency of D3-instantons with S-duality was
established at first order in the instanton expansion, using the modular properties of the M5-brane elliptic genus.
We extend this analysis to the two-instanton level, where wall-crossing phenomena start playing a role.
We focus on the contact potential, an analogue of the K\"ahler potential which must transform as
a modular form under S-duality. We show that it can be expressed in terms of
a suitable modification of the partition function of D4-D2-D0 BPS black holes,
constructed out of the generating function of MSW invariants
(the latter coincide with Donaldson-Thomas invariants in a particular chamber).
Modular invariance of the contact potential
then requires that, in case where the D3-brane wraps a reducible divisor,
the generating function of MSW invariants must transform as a vector-valued mock modular form,
with a specific modular completion built from the MSW invariants of the constituents.
Physically, this gives a powerful constraint on the degeneracies of BPS black holes.
Mathematically, our result gives a universal prediction for the modular properties of
Donaldson-Thomas invariants of pure two-dimensional sheaves.
}

\begin{document}

\section{Introduction}

The low energy effective action of type II string theory compactified on a Calabi-Yau threefold is determined by the metric
on the moduli space, which is a direct product of its vector multiplet and hypermultiplet components.
Whereas the former is classically exact, the hypermultiplet moduli space $\cM_H$ receives a variety of quantum corrections
(see e.g. \cite{Alexandrov:2011va,Alexandrov:2013yva} and references therein).
In type IIB string theory, if the volume of the Calabi-Yau threefold $\CYm$ is taken to be large in string units,
these quantum corrections can be ordered according to the following hierarchy:
i) one-loop and D(-1) instanton corrections,
ii) $(p,q)$ string instantons,
iii) D3-instantons,
iv) $(p,q)$ five-brane instantons.
All these corrections are expected to be governed by topological invariants of $\CYm$,
including its intersection form $\kappa_{abc}$, Euler number $\chi_{\CYm}$,
Chern classes $c_{2,a}$, genus zero Gromov-Witten invariants $n_{q_a}$ and Donaldson-Thomas (DT) invariants
$\Omega(\gamma;z^a)$.\footnote{Here the index $a$ runs over $1,\dots,b_2(\CYm)$,
$q_a$ labels effective homology classes $H_2^+(\CYm)$, $\gamma$ labels vectors in
the homology lattice $H^{\rm even}(\CYm)$, and $z^a=b^a+\I t^a$ are complexified K\"ahler moduli.}
In addition, they are severely constrained by the fact that the exact metric
on $\cM_H$ should be \qk \cite{Bagger:1983tt} and smooth across walls of marginal stability
\cite{Gaiotto:2008cd,Alexandrov:2008gh} in spite of the discontinuities of the DT invariants
$\Omega(\gamma;z^a)$.
Most notably, it should carry an isometric action of the modular
group $SL(2,\IZ)$ \cite{RoblesLlana:2006is},
originating from the S-duality symmetry in uncompactified type IIB string theory.

In order to satisfy the first requirement, it is most convenient to use the twistorial
formulation of \qk manifolds  \cite{MR1327157,Alexandrov:2008nk}. In this framework,
quantum corrections to the metric on $\cM_H$ are captured by a set of holomorphic
functions on the twistor space $\cZ$ of $\cM_H$, which encode gluing conditions between
local Darboux coordinate systems for the canonical complex contact structure on $\cZ$.
Furthermore, discrete isometries of $\cM_H$ must  lift to holomorphic coordinate transformations
on $\cZ$ preserving the contact structure, which constrains the possible gluing conditions.
In the presence of a continuous isometry,  another important object, central
for this work, is the contact potential $e^\Phi$, a real function on $\cM_H$,  defined as the norm
of the moment map for the corresponding isometry \cite{Alexandrov:2008nk,Alexandrov:2011ac}.
Its importance lies in the fact that it provides a \kahler potential on $\cZ$, and that
it must be invariant under any further discrete isometry, up to a rescaling dictated by
the transformation of the contact one-form.
In the present context, the isometry corresponds to translation along the NS axion,
which is broken only by $(p,q)$ five-brane instantons, while the contact potential
determines the 4-dimensional string coupling.

Since the action of the modular group preserves the large volume limit, modular invariance
should hold at each level in the aforementioned hierarchy of quantum corrections.
For the first two levels, modular invariance was used in \cite{RoblesLlana:2006is} to
infer the D(-1) and $(p,q)$-string instanton corrections
from the known  world-sheet instantons at tree-level and the one-loop correction.
The contributions of D3 and D5 instantons were then deduced by requiring symplectic invariance
and smoothness across walls of marginal stability \cite{Alexandrov:2008gh,Alexandrov:2009zh}.
The consistency of D3-instantons with S-duality
however  depends on special properties
of the DT invariants $\Omega(\gamma;z^a)$, where in this case $\gamma$ labels
the charges $(p^a, q_a,q_0)$ of a D3-D1-D(-1) instanton, or more mathematically,
the Chern character of a coherent sheaf with support on an effective divisor $\cD$ in $\CYm$.

In order to study this problem, it is useful to express the DT invariants $\Omega(\gamma;z^a)$,
which in general exhibit wall-crossing behavior with respect to the \kahler moduli $z^a$,
in terms of the so-called Maldacena-Strominger-Witten (MSW) invariants $\OmMSW(\gamma)$,
familiar from the study of the partition function of D4-D2-D0 black holes
\cite{Maldacena:1997de}.
Unlike DT invariants, MSW invariants are independent of the moduli.
Moreover, in the case where the divisor $\cD$ wrapped by the D4-brane is irreducible
(in the sense that $\cD$ cannot be written as the sum of two effective divisors)\footnote{This irreducibility condition
has not been fully appreciated in the past, and part of the present work aims at relaxing it.},
the MSW invariants appear as Fourier coefficients of a modular form, namely the elliptic genus
$\chi_\bfp(\tau,z^a,c^a)$ of the superconformal field theory describing an M5-brane
wrapped on $T^2\times \cD$ \cite{Maldacena:1997de}.
More precisely, the elliptic genus decomposes into\footnote{\label{footJacobi}
The elliptic genus is usually a function of the modular parameter $\tau$ and of complex
parameters $v^a\in \mathbb{C}$ coupling to conserved currents in the SCFT, and transforms
as a Jacobi form of fixed weight and index. In contrast, the function defined in \eqref{defZ1intro}
depends on the \kahler moduli $z^a=b^a+\I t^a$ and RR potentials $c^a$ at spatial infinity,
which decouple in the near-horizon geometry, and transforms as an ordinary modular form
of weight $(-\tfrac32, \tfrac12)$.
The standard elliptic genus is obtained by
specializing $t^a$ to the large volume attractor point $\lambda p^a$ with $\lambda\to +\infty$, 
setting $b^a=0$ and analytically continuing in $c^a$. With this understanding, we shall
nonetheless refer to \eqref{defZ1intro} as the elliptic genus of the M5-brane SCFT. Incidentally,
we warn the reader that our definition of theta series
is complex conjugate of the usual one used in \cite{Alexandrov:2012au}.
This avoids a proliferation of complex conjugations
and facilitates comparison with the results of the twistorial
formalism.}
\be
\label{defZ1intro}
\chi_\bfp(\tau,z^a,c^a)= \sum_{\bfmu\in\Lambda^\star/\Lambda}
h_{\bfp,\bfmu}(\tau)\,\theta_{\bfp,\bfmu}(\tau,t^a,b^a,c^a),
\ee
where $\theta_{\bfp,\bfmu}$ is the Siegel theta series \eqref{defth},
a vector-valued modular form of weight $(\frac{b_2(\CYm)-1}{2},\tfrac12)$, and $h_{\bfp,\bfmu}$ is the
generating function \eqref{defhmu} of MSW invariants. When $\cD$ is irreducible,
$h_{\bfp,\bfmu}$ is a holomorphic vector-valued modular form of weight $(-\frac{b_2(\CYm)}{2}-1,0)$, so that
$\chi_\bfp(\tau,z^a,c^a)$ transforms as  a modular form of
weight $(-\tfrac32, \tfrac12)$, as expected from the elliptic genus of a standard SCFT
\cite{Gaiotto:2006wm,deBoer:2006vg,Denef:2007vg}.

DT invariants coincide with  MSW invariants at the `large volume attractor point',
but in general receive additional contributions proportional to
products of MSW invariants with moduli-dependent coefficients,
corresponding to black hole bound states \cite{Manschot:2009ia, Manschot:2010xp}. The
D3-instanton corrections to the metric can thus be organized as an
infinite series in powers of MSW invariants, corresponding to multi-instanton effects.
In \cite{Alexandrov:2012au} we considered the one-instanton approximation (and large volume limit)
of the D3-instanton corrected metric on $\cM_H$, keeping only the first term of the expansion \eqref{multiMSW} of DT
invariants in terms of MSW  invariants.
Relying on the modular properties of MSW invariants encoded in the elliptic genus \eqref{defZ1intro},
we showed that in this approximation, the metric on $\cM_H$ admits an isometric action of the modular group.
This result was achieved by showing that S-duality acts on the  canonical Darboux coordinates
on $\cZ$ introduced in \cite{Alexandrov:2008gh,Alexandrov:2009zh} by a holomorphic contact transformation.
While the transformation properties of Darboux coordinates are, already at the classical level,
quite complicated, S-duality requires that the contact potential $e^\Phi$
should transform in a simple way, namely as a modular form of weight $(-\tfrac12,-\tfrac12)$.
In \cite{Alexandrov:2012au} we proved that this
is the case by showing that the contact potential is directly related to the elliptic genus \eqref{defZ1intro}
via the action of a modular covariant derivative.

In this paper, we study the corrections to the metric on $\cM_H$ at the two-instanton level,
i.e. at order $(\OmMSW)^2$ in the expansion in powers of MSW invariants. The analysis of
the transformation properties of Darboux coordinates and a complete proof of the existence of
an isometric action of S-duality on $\cM_H$
is deferred to  a subsequent paper \cite{abmp-to-appear}. In this paper, we shall restrict our attention
to the contact potential, which is much simpler but yet encodes all possible quantum corrections.

At two-instanton order, we must take into account both corrections to the contact potential
which are quadratic in the DT invariants,
and order $(\OmMSW)^2$ contributions in the relation between DT and MSW invariants.
Our main result is as follows: the contact potential can be expressed in terms of the modular
covariant derivative of the following BPS partition function
\be
\label{defZthintro}
\hcZ_\bfp=
\sum_{\bfmu\in\Lambda^\star/\Lambda}
\whh_{\bfp,\bfmu}\,\theta_{\bfp,\bfmu}
+ \frac{1}{2} \sum_{\bfp_1+\bfp_2=\bfp}\sum_{\bfmu_i\in \Lambda^\star/\Lambda_i}
\whh_{\bfp_1,\bfmu_1}\,\whh_{\bfp_2,\bfmu_2}\, \widehat \Psi_{\bfp_1,\bfp_2,\bfmu_1,\bfmu_2}
+\cdots ,
\ee
where the dots denote terms of higher order in $\whh_{\bfp,\bfmu}$.
Here two new objects are introduced:
\begin{itemize}
\item
$\widehat \Psi_{\bfp_1,\bfp_2,\bfmu_1,\bfmu_2}$ is the non-holomorphic theta series constructed in \cite{Manschot:2009ia}
for the lattice of signature $(2,2b_2-2)$ spanned by the D1-brane
charges $(q_1,q_2)$ of the two constituents. It transforms as a vector valued modular form of weight
$(b_2(\CYm)+\tfrac12,\tfrac12)$ and captures the wall-crossing dependence of $\hcZ_\bfp$ due to two-center black
hole solutions (or equivalently two-centered D3-instantons).
\item
$\whh_{\bfp,\bfmu}=h_{\bfp,\bfmu}-\hf\,R_{\bfp,\bfmu}$,
where $R_{\bfp,\bfmu}$ is a non-holomorphic function of $\tau$
constructed out
of the MSW invariants,
\bea
R_{\bfp,\bfmu}(\tau)&=&-\frac{1}{4\pi} \sum_{\bfp_1+\bfp_2=\bfp}
\sum_{\bfmu_i\in \Lambda^\star/\Lambda_i}
h_{\bfp_1,\bfmu_1}(\tau)\,h_{\bfp_2,\bfmu_2}(\tau)
\sum_{\bfrho\in (\Lambda_1-\tilde \mu) \cap (\Lambda_2+\tilde \mu)}
(-1)^{\Sgg}
\nn\\
&&\times
\left|\Sgg\right|
\beta_{\frac{3}{2}}\!\left(\textstyle{\frac{2\tau_2 \(\Sgg\)^2 }{(pp_1p_2)}}\right)
e^{\pi\I \tau {\rm Q}_{\bfp_1,\bfp_2}\!\!\left(\bfnu_1,\bfnu_2\right)},
\label{defRmu}
\eea
where $\beta_{\frac{3}{2}}$ is the function defined in \eqref{betahaf} and the definitions of other notations
can be found in Appendix \ref{ap-R}.
\end{itemize}
When the effective divisor $\cD$ is irreducible, the sum over $\bfp_1,\bfp_2$ is empty so that $R_{\bfp,\bfmu}$
and the second term in \eqref{defZthintro} vanish and $\hcZ_\bfp$ reduces to the elliptic genus \eqref{defZ1intro}.
If on the contrary $\cD$ can be decomposed into a sum of two effective divisors $\cD_1+\cD_2$, then modular invariance of the contact
potential requires that the non-holomorphic function $\whh_{\bfp,\bfmu}$
must transform as a (vector-valued) modular form of weight $(-\frac{b_2}{2}-1,0)$.
This shows that the holomorphic generating function  $h_{\bfp,\bfmu}$
is not a modular form, but rather a (mixed) mock modular form \cite{Zwegers-thesis, MR2605321}.

A similar modular anomaly was in fact observed
long ago for the partition function of topologically twisted $\cN=4$
Yang-Mills theory with gauge group $U(2)$ on a complex surface in \cite{Vafa:1994tf}
and, more recently, in \cite{Manschot:2011dj}. This set-up was related to the case
of multiple M5-branes wrapped on a rigid divisor in a non-compact
threefold in \cite{Minahan:1998vr, Alim:2010cf}. For M5-branes wrapped on non-rigid
divisors in an elliptically fibered compact threefold, such an anomaly was also argued to
appear in \cite{Klemm:2012sx} using the holomorphic anomaly in topological
string theory \cite{Bershadsky:1993ta} and T-duality. However in the
latter context the anomaly is of quasi-modular type rather then mock-modular.

Modular or holomorphic anomalies are also
known to occur in the context of quantum gravity partition functions
for AdS$_3$/CFT$_2$ \cite{Manschot:2007ha}, non-compact coset conformal field theories \cite{Troost:2010ud},
and partition functions for BPS black holes in $\cN=4$
supergravity \cite{Dabholkar:2012nd}. In the context of black hole partition functions,
the non-holomorphic completion was related to the spectral anomaly in the continuum of scattering states
in \cite{Pioline:2015wza}. Our result shows that modular or holomorphic
anomalies generally affect M5-branes or
D4-branes wrapped on reducible divisors in an arbitrary compact Calabi-Yau threefold,
and gives a precise prediction for the modular completion in the case where $\cD$ is
the sum of two  irreducible divisors.
Physically, this gives a powerful constraint on the degeneracies of D4-D2-D0
brane black holes composed of two D4-branes. In particular, the mock modularity of $h_{\bfp,\bfmu}$
affects the asymptotic growth of the degeneracies \cite{Bringmann:2010sd}.
Mathematically, upon re-expressing the MSW invariants in terms of DT-invariants, our result gives a universal prediction
for the modularity  of  DT invariants for pure 2-dimensional
sheaves, which is receiving increasing attention from the mathematics
community,  see e.g.  \cite{Toda:2013, Toda:2014, Gholampour:2013hfa, Diaconescu:2015koa, Bouchard:2016lfg}. Using similar techniques,
it should be possible in principle to determine the modular anomaly in the case
where $\cD$ can split into a sum of more than two irreducible divisors.

The organization of the paper is as follows.
In section \ref{sec-BPS} we discuss the BPS invariants counting D3-brane instantons and associated modular forms.
In section \ref{sec-twist}, we review the twistorial formulation of the D-instanton corrected hypermultiplet moduli space
of type IIB string theory compactified on a Calabi-Yau threefold.
Then in section \ref{sec-contact} we compute the D3-instanton contribution to the contact potential in
the two-instanton approximation and express it in terms of $\hcZ_\bfp$.
Finally, we conclude in section \ref{sec-concl}.
Appendices \ref{ap-indef}, \ref{ap-R} and \ref{ap-cp} contain some useful material and details of our calculations.

\section{BPS invariants for D3-instantons and mock modularity}
\label{sec-BPS}

In this section, we discuss the modular properties of the BPS
invariants which control D3-brane instanton corrections
to the hypermultiplet moduli space $\cM_H$ in type IIB string theory compactified on a Calabi-Yau threefold $\CYm$.
The same invariants also control the degeneracies of D4-D2-D0 black holes in type IIA string theory compactified
on the same threefold $\CYm$. When the D3-brane wraps a primitive
effective divisor\footnote{We will identify a divisor $\cD$ with its
  class in $H_4(\CYm,\mathbb{Z})$. We call a divisor $\cD$ {\it irreducible}
  if it is an irreducible analytic hypersurface of $\CYm$
  \cite{MR95d:14001}. Let $\gamma_a$ be a
  set of $b_2$ irreducible divisors forming a basis of
  $H_4(\CYm,\mathbb{Z})$. Then a divisor $\cD=\sum_a
  r^a \gamma_a$ is {\it effective} if $r^a\geq 0$ for all
$a$, and not all equal to $0$ simultaneously. We call a divisor {\it primitive} if gcd$(\{r^a\})$=1.
}
$\cD$, these invariants are claimed to be Fourier coefficients
of a vector-valued modular form. Instead, we will argue that, when $\cD=\sum_{i=1}^n \cD_i$
is the sum of $n$ irreducible divisors, the invariants are the coefficients of
the {\it holomorphic part} of a real-analytic modular form. For $n=2$, we show that this holomorphic part is in fact
a {\it mixed mock} modular form, whose modular anomaly is controlled by the invariants associated to $\cD_i$.

\subsection{D3-instantons, DT and MSW invariants}

Let us first introduce some mathematical objects and notations relevant for D3-instantons.
As in \cite{Alexandrov:2012au}, we denote by
$\gamma_a$ an integer irreducible basis of $ \Lambda=H_4(\CYm,\IZ)$, $\omega_a$ their Poincar\'e dual 2-forms, $\gamma^a$
an integer basis of $\Lambda^*=H_2(\CYm,\IZ)$, $\omega^a$ their Poincar\'e dual 4-forms,
and $\omega_\CYm$ the volume form of $\CYm$ such that
\be
\omega_a \wedge\omega_b = \kappa_{abc}\, \omega^c ,
\qquad
\omega_a \wedge \omega^b = \delta_a^b \, \omega_\CYm ,
\qquad
 \int_{\gamma^a}\omega_b= \int_{\gamma_b}\omega^a= \delta^a_b  ,
\ee
where $\kappa_{abc}$ is the intersection form, integer-valued and symmetric in its indices.
For brevity we shall denote $(lkp)=\kappa_{abc}l^a k^b p^c$ and
$(kp)_a=\kappa_{abc}k ^b p^c$.

A D3-instanton is described by a coherent sheaf $\cE$ of rank $r$ supported on a divisor
$\cD \subset \CYm$. The homology class of the divisor $\cD$ may be expanded on the
basis of 4-cycles as $\cD= d^a \gamma_a$. We assume that $\cD$ is
effective, and furthermore that its Poincar\'e dual $[\cD]$ belongs to
the \kahler cone,
\be
\label{khcone}
d^3> 0,
\qquad
(r d^2)> 0,
\qquad
k_a d^a > 0,
\ee
for all effective divisors $r^a \gamma_a \in H_4^+(\CYm,\IZ)$ and
effective curves $k_a \gamma^a \in H_2^+(\CYm,\IZ)$. We expect
however that our results can be generalized to cases where $[\cD]$ lies
on the boundary of the \kahler cone.

The D-brane charges are given
by the components of the generalized Mukai vector of $\cE$ on a basis of $H^{\rm even}(\CYm,\IZ)$,
\be
\gamma= \ch \cE \, \sqrt{\Td \CYm}
= p^a \omega_a - q_a \omega^a + q_0\, \omega_{\CYm}\, ,
\ee
where $p^a = r d^a$. The charges $p^a,q_a,q_0$
satisfy the following quantization conditions\footnote{The electric charges $q_a$ and $q_0$ (denoted by $q'_a, q'_0$
in \cite{Alexandrov:2010ca}) are not integer valued. They are related to the integer
charges which appear naturally on the type IIA side by a rational symplectic transformation \cite{Alexandrov:2010ca}.
\label{foot-charges}}
\be
\label{fractionalshiftsD5}
p^a\in\IZ ,
\qquad
q_a \in \IZ  + \frac12 \,\kappa_{abc} p^b p^c ,
\qquad
q_0\in \IZ-\frac{1}{24}\, p^a c_{2,a} .
\ee
We denote the corresponding charge lattice by $\Gamma$, and its intersection
with the \kahler cone \eqref{khcone} by $\Gamma_+$.
Upon tensoring the sheaf $\cE$ with a line bundle $\cL$ on $\cD$, with $c_1(\cL)=-\epsilon^a \omega_a$,
the magnetic charge $p^a$ is invariant, while the electric charges $q_a, q_0$
vary by a `spectral flow'
\be
\label{flow}
q_a \mapsto q_a - \kappa_{abc} p^b \epsilon^c,
\qquad
q_0 \mapsto q_0 - \epsilon^a q_a + \frac12\, \kappa_{abc} p^a \epsilon^b \epsilon^c.
\ee
This transformation leaves invariant the combination
\be
\label{defqhat}
\hat q_0 \equiv
q_0 -\frac12\, \kappa^{ab} q_a q_b ,
\ee
where $\kappa^{ab}$ is the inverse of $\kappa_{ab}=\kappa_{abc} p^c$, a quadratic form
of signature $(1,b_2-1)$ on $\Lambda\otimes \IR\simeq \IR^{b_2}$.
We use this quadratic form to identify $\Lambda\otimes \IR$ and $\Lambda^*\otimes \IR$,
and use bold-case letters to denote the corresponding vectors. We also
identify $\Lambda$ with its image in $\Lambda^*$. Note however that the map $\epsilon^a \mapsto \kappa_{ab} \epsilon^b$
is in general not surjective:
the quotient $\Lambda^*/\Lambda$ is a finite group of order $|\det\kappa_{ab}|$.
The transformation \eqref{flow} preserves the
residue class $\mu_a\in \Lambda^*/\Lambda$ defined by
\be
\label{defmu}
q_a = \mu_a + \frac12\, \kappa_{abc} p^b p^c + \kappa_{abc} p^b \epsilon^c,
\qquad
\bfeps\in\Lambda\, .
\ee
We note also that the invariant charge $\hat q_0$ is bounded from above by $\hat q_0^{\rm max}=\tfrac{1}{24}(p^3+c_{2,a}p^a)$.

The contribution of a single D3-instanton to the metric on $\cM_H$ is proportional to the DT invariant
$\Omega(\gamma;\bfz)$, which is  the (weighted) Euler characteristic of the moduli
space of semi-stable sheaves with fixed Mukai vector $\gamma$. The relevant stability
condition is $\Pi$-stability \cite{Douglas:2000gi}, which reduces to slope stability
in the large volume limit. The latter stability condition states that
for each subsheaf $\cE'(\gamma')\subset \cE(\gamma)$ the following inequality is satisfied
\be
\label{stability}
\frac{(q'_a+(bp')_a)t^a}{(p't^2)}\leq \frac{(q_a+(bp)_a)t^a}{(pt^2)}.
\ee

It is useful to define the rational DT invariant \cite{ks,Joyce:2008pc,Manschot:2010qz},
\be
\label{defntilde}
\bar\Omega(\gamma;\bfz) = \sum_{d|\gamma}  \frac{1}{d^2}\,
\Omega(\gamma/d;\bfz)\, ,
\ee
which reduces to the integer-valued DT invariant $\Omega(\gamma;\bfz)$ when
$\gamma$ is a primitive vector, but is in general rational-valued. Both $\Omega$
and $\bar\Omega$ are
piecewise constant as a function of the complexified K\"ahler moduli $z^a=b^a+\I t^a$,
but are discontinuous across walls of marginal stability where the
sheaf becomes unstable, i.e. the codimension-one subspaces of the
K\"ahler cone across which the inequality \eqref{stability} flips.
$\Omega$ and $\bar\Omega$ are in general {\it not} invariant
under the spectral flow \eqref{flow}, but they are invariant under the combination of
\eqref{flow} with a compensating shift of the Kalb-Ramond field, $b^a\mapsto b^a +\epsilon^a$.

A physical way to understand the moduli dependence of $\Omega(\gamma;\bfz)$ is to note
that the same invariant counts D4-D2-D0 brane bound states in type IIA theory compactified
on the same CY threefold $\CYm$. The mass of a single-particle BPS state is equal to the modulus of
the central charge function $Z_\gamma = q_\Lambda z^\Lambda - p^\Lambda F_\Lambda(z)$
(where $\Lambda=(0,a)=0,\dots,b_2$ and $F_\Lambda=\p_{X^\Lambda}F(X)$
is the derivative of the holomorphic prepotential $F$).
Some of these single-particle BPS states may however arise as bound states of more elementary
constituents with charge $\gamma_i$ such that $\sum_i\gamma_i=\gamma$. Typically,
these bound states exist only in some chamber in \kahler moduli space, and decay across
walls of marginal stability where the central charges $Z(\gamma_i)$ become aligned,
so that the mass $|Z_\gamma|$ coincides with the sum $\sum_i |Z_{\gamma_i}|$ of the masses of the constituents.
A similar picture exists for D3-instantons, where the modulus of the central charge controls
the classical action, but the analogue of the notion of single-particle state is somewhat obscure.

At the special value of the moduli
$\bfz(\gamma)$ given by the attractor mechanism \cite{Ferrara:1995ih}, no bound states exist,
and therefore $\Omega(\gamma;\bfz(\gamma))$
counts elementary states, which cannot decay. Since we are
only interested in the large volume limit, we  define the `MSW invariants'
$\OmMSW(\gamma)=\Omega(\gamma;\bfz_\infty(\gamma))$ as the DT invariants evaluated at the large volume attractor point,
\be
\label{lvolatt}
\bfz_\infty(\gamma)=\lim_{\lambda\to +\infty}\(\bfb(\gamma)+\I\lambda \bft(\gamma)\)
= \lim_{\lambda\to +\infty}\(-\bfq+\I\lambda  \bfp\).
\ee
The reason for the name MSW (Maldacena-Strominger-Witten) is that
when $\bfp$ corresponds to a very ample primitive divisor, these states are in fact described by the
superconformal field theory discussed in \cite{Maldacena:1997de}.
It is important that, due to the symmetry \eqref{flow}, $\OmMSW(\gamma)$ only
depend on $p^a, \mu_a$ and $\hat q_{0}$ defined in \eqref{defqhat} and \eqref{defmu}.
We shall therefore write $\OmMSW(\gamma)=\OmMSW_{\bfp,\bfmu}( \hat q_0)$.

Away from the large volume attractor point (but still in the large volume limit),
the DT invariant $\Omega(\gamma;\bfz)$ receives additional contributions from bound states with charges
$\gamma_i=(0,p_i^a,q_{i,a},q_{i,0})\in\Gamma_+$ such that $\sum_i
\gamma_i=\gamma$ and $\bfp_i\neq 0$ for each $i$.
For $n=2$, the case of primary interest in this work, bound states exist if and only if the sign
of ${\rm Im}(Z_{\gamma_1} \bar Z_{\gamma_2})$ is equal to the sign of
$\langle\gamma_1,\gamma_2 \rangle=p^\Lambda_2q_{1,\Lambda}-p^\Lambda_1 q_{2,\Lambda}$
\cite{Denef:2000nb}. In the large volume limit, one has
\be
{\rm Im}(Z_{\gamma_1} \bar Z_{\gamma_2})  =
-\frac12 \sqrt{(p_1t^2)\, (p_2 t^2) \, (pt^2)}\, \cI_{\gamma_1\gamma_2},
\ee
where\footnote{\label{foobind}It is worth recognizing that $\cI_{\gamma_1\gamma_2}^2$ is equal 
to the large volume limit of the binding energy
$|Z_{\gamma_1}|+|Z_{\gamma_2}|-|Z_{\gamma_1+\gamma_2}|$ (as follows from 
\eqref{ZlargeV} and \eqref{IZid}). In particular, it vanishes on the wall of marginal stability.}
\be
\cI_{\gamma_1\gamma_2}= \frac{(p_2 t^2)\(q_{1,a}+(bp_1)_a\)t^a- (p_1 t^2)\(q_{2,a}+(bp_2)_a\)t^a}
{\sqrt{(p_1t^2)\, (p_2 t^2) \, (pt^2)}}
\label{defIgg}
\ee
is invariant under rescaling of $t^a$.
It is convenient to define the `sign factor'
\be
\Dela_{\gamma_1\gamma_2}=
\hf\,\Bigl(\sgn\!\left(\cI_{\gamma_1\gamma_2}(\bft)\right)-\sgn\!\left(\langle\gamma_1,\gamma_2 \rangle\right)\Bigr),
\label{defDela}
\ee
where we indicated explicitly the dependence on the K\"ahler moduli.
This factor takes the value $\pm 1$ when bound states are allowed, or $0$ otherwise.
The DT invariants are then expressed in terms of the MSW invariants by
\cite{Manschot:2009ia}
\bea
\bar\Omega(\gamma;\bfz)&=&\bOmMSW(\gamma)
+\frac{1}{2}\sum_{\gamma_1,\gamma_2\in \Gamma_+\atop \gamma_1+\gamma_2=\gamma}
(-1)^{\langle\gamma_1,\gamma_2\rangle} \langle\gamma_1,\gamma_2\rangle
\,\Dela_{\gamma_1\gamma_2}\,
\bOmMSW(\gamma_1)\, \bOmMSW(\gamma_2)
+\cdots,
\label{multiMSW}
\eea
where the dots denote contributions of higher order in the MSW invariants.

\subsection{Modularity of the BPS partition function}
\label{subsec-modul}

Let us now consider the partition function of DT invariants with fixed
magnetic charge $\bfp$. Let $\tau=\tau_1+i\tau_2\in
\mathbb{H}$, $\bfc\in \mathbb{R}^{b_2}$ the RR potentials conjugate to
D1-brane charges, and $\bfb\in \mathbb{R}^{b_2}$ the Kalb-Ramond
field. The BPS partition function is defined as the following generating function of
DT-invariants
\be
\label{defZp}
\mathcal{Z}_{\bfp}(\tau,{\bfz}, \bfc)=e^{\pi \tau_2(pt^2)}
\sum_{q_\Lambda}
\bar\Omega(\gamma;\bfz)\, (-1)^{\bfp\cdot\bfq}\, e^{-2\pi \tau_2 |Z_\gamma|-2 \pi \I \tau_1
\(q_0+\bfb\cdot \bfq+\half \bfb^2\)+2\pi \I \bfc\cdot \(\bfq+\half\bfb\)},
\ee
where the sum goes over charges satisfying the quantization conditions
\eqref{fractionalshiftsD5}. The DT-invariants are weighted by the Boltzmann factor $\exp(-2\pi
\tau_2|Z_\gamma|)$ and by a phase factor induced by  the couplings of the charges to
the potentials $\tau_1$, $\bfb$ and $\bfc$.
The factor $(-1)^{\bfp\cdot \bfq}$ is motivated by modular properties of $\mathcal{Z}_{\bfp}$, whereas
the prefactor $e^{\pi \tau_2(pt^2)}$ is included so as to subtract the leading divergent term
in the large volume limit of $|Z_\gamma|$:
\be
\label{ZlargeV}
|Z_\gamma|=\half (p t^2)-q_0+(q+b)_+^2-(\bfq+\half\bfb)\cdot\bfb +\cdots.
\ee
Here the dots denote terms of order $1/(p t^2)$ and, as in \cite{Alexandrov:2012au}, we defined
\be
\bfq_+ = \frac{q_a t^a}{(p t^2)}\, \bft\, ,
\qquad
\bfq_- = \bfq - \bfq_+\, ,
\qquad
q_+ = \frac{q_a t^a}{\sqrt{(p t^2)}}\, ,
\ee
so that $q_+^2=(\bfq_+)^2=\bfq^2 - (\bfq_-)^2$.
In the following we shall study the behavior of the BPS partition function \eqref{defZp}
under modular transformations.

Substituting \eqref{multiMSW} into \eqref{defZp}, one obtains an expansion in powers of the MSW invariants
\be
\mathcal{Z}_{\bfp}(\tau,{\bfz}, \bfc) = \sum_{n\geq 1} \mathcal{Z}^{(n)}_{\bfp}(\tau,{\bfz}, \bfc),
\ee
where $\mathcal{Z}^{(n)}_{\bfp}$ corresponds to the terms of degree $n$ in
$\bOmMSW(\gamma_i)$. Due to the symmetry of the MSW invariants under the spectral flow \eqref{flow},
all terms in this expansion have a theta series decomposition. Indeed,
decomposing the vectors $\bfq_i$ according to \eqref{defmu},
we find, for the first \cite{Gaiotto:2006wm,deBoer:2006vg,Denef:2007vg}
and second \cite{Manschot:2009ia} terms
\be
\label{defZ1}
\mathcal{Z}^{(1)}_\bfp(\tau,\bfz,\bfc)= \chi_{\bfp}(\tau,\bfz,\bfc)=\sum_{\bfmu\in\Lambda^\star/\Lambda}
h_{\bfp,\bfmu}(\tau)\,\theta_{\bfp,\bfmu}(\tau,\bft,\bfb,\bfc),
\ee
\be
\mathcal{Z}^{(2)}_\bfp(\tau,\bfz,\bfc)=\frac{1}{2} \sum_{ \bfp_1+\bfp_2=\bfp}
\sum_{\bfmu_i\in \Lambda^\star/\Lambda_i}
h_{\bfp_1,\bfmu_1}(\tau)\,h_{\bfp_2,\bfmu_2}(\tau)\, \Psi_{\bfp_1,\bfp_2,\bfmu_1,\bfmu_2}(\tau,\bft,\bfb,\bfc) .
\label{defZ2}
\ee
Here, we denote by $\Lambda_i$ the image of $\Lambda$ inside $\Lambda^*$ under the map $\eps^a\mapsto \kappa_{abc}\eps^b p_i^c$
and introduce the following objects (we denote $\expe{x}=e^{2\pi\I x}$):
\begin{itemize}
\item
a holomorphic function of the modular parameter $\tau$ built from the MSW invariants
\be
\label{defhmu}
h_{\bfp,\bfmu}(\tau) = \sum_{\hat q_0\le \hat q_0^{\rm max}}
\bOmMSW_{\bfp,\bfmu}(\hat q_0)\,
\expe{-\hat q_0 \tau };
\ee
\item
the Siegel-Narain theta series
\be
\label{defth}
\theta_{\bfp,\bfmu}(\tau,\bft,\bfb,\bfc)=\sum_{\bfk\in\Lambda+\bfmu+\half \bfp} (-1)^{\bfk\cdot \bfp}\,\cXt_{\bfp,\bfk}\, ,
\ee
where
\be
\cXt_{\bfp,\bfk}=\expe{ - \frac{\tau}{2}\,(\bfk+ \bfb)_-^2-\frac{\bar\tau}{2}\, (\bfk+ \bfb)_+^2 +\bfc \cdot(\bfk +\haf \bfb)};
\label{defXt}
\ee
\item
the `mock Siegel-Narain theta series' which is a sum over the double
lattice $\Lambda_1\oplus \Lambda_2$\cite{Manschot:2009ia}
\be
\label{defPsi}
\Psi_{\bfp_1,\bfp_2,\bfmu_1,\bfmu_2}(\tau,\bft,\bfb,\bfc)=
\!\!\sum\limits_{\bfk_i \in \Lambda_i+\bfmu_i+\half\bfp_i}\!\!
(-1)^{\bfp_1\cdot\bfk_1+\bfp_2\cdot\bfk_2+(p_1^2p_2)}
\langle\gamma_1,\gamma_2\rangle\,\Dela_{\gamma_1\gamma_2}
e^{2\pi\tau_2\cI_{\gamma_1\gamma_2}^2}\cXt_{\bfp_1,\bfk_1}\cXt_{\bfp_2,\bfk_2} ,
\ee
where $\cI_{\gamma_1\gamma_2}$ and $\Dela_{\gamma_1\gamma_2}$ are defined in
\eqref{defIgg} and \eqref{defDela}.
\end{itemize}

The theta series decompositions \eqref{defZ1} and \eqref{defZ2}
provide the starting point to discuss the modular properties of
the BPS partition function. The modular group acts by the following transformations
\be\label{SL2Z}
\begin{split}
&\quad \tau \mapsto \frac{a \tau +b}{c \tau + d} \, ,
\qquad
\bft \mapsto |c\tau+d| \,\bft,
\qquad
\begin{pmatrix} \bfc \\ \bfb \end{pmatrix} \mapsto
\begin{pmatrix} a & b \\ c & d  \end{pmatrix}
\begin{pmatrix} \bfc \\ \bfb \end{pmatrix} ,
\end{split}
\ee
with $ad-bc=1$. Under this action,  the theta series $\theta_{\bfp,\bfmu}$
is well-known to transform as a vector-valued modular form of weight
$(\tfrac{b_2-1}{2},\tfrac12)$ and multiplier system $M_\theta$.
In contrast, the double theta series \eqref{defPsi} does {\it not} transform as a
vector-valued modular form under $SL(2,\IZ)$. However, it was shown in \cite{Manschot:2009ia},
using similar techniques as in \cite{Zwegers-thesis}, that it can be completed into
a vector-valued modular form $\widehat\Psi=\Psi+\Psi^{(+)}+\Psi^{(-)}$
of weight $( b_2+\tfrac12,\tfrac12)$, at the expense
of adding two double theta series of the form
\be
\Psi^{(\pm)}_{\bfp_1,\bfp_2,\bfmu_1,\bfmu_2}(\tau,\bft,\bfb,\bfc)=
\sum\limits_{\bfk_i \in \Lambda_i+\bfmu_i+\half\bfp_i}
(-1)^{\bfp_1\cdot\bfk_1+\bfp_2\cdot\bfk_2+(p_1^2p_2)}\,\Pi^{(\pm)}_{\gamma_1\gamma_2}
\,e^{2\pi\tau_2\cI_{\gamma_1\gamma_2}^2}\cXt_{\bfp_1,\bfk_1}\cXt_{\bfp_2,\bfk_2},
\label{defPsih}
\ee
where the insertions are given by the following expressions
\bea
\Pi^{(+)}_{\gamma_1\gamma_2}&=&{\sqrt{\frac{(p t^2)\,(p_1p_2 t)^2}{8\pi^2\tau_2\,(p_1t^2)\,(p_2t^2)}}}\,
e^{-2\pi \tau_2\cI_{\gamma_1,\gamma_2}^2}
-\hf\,\langle\gamma_1,\gamma_2\rangle\, \sgn(\cI_{\gamma_1\gamma_2})\,
\beta_{\half}\!\left(2\tau_2\cI_{\gamma_1\gamma_2}^2 \right) ,
\label{ins-one} \\
\Pi^{(-)}_{\gamma_1\gamma_2}
&=&-\frac{1}{4\pi}
\abs{\langle\gamma_1,\gamma_2\rangle} \,
\beta_{\frac{3}{2}}\!\left({\frac{2\tau_2\langle\gamma_1,\gamma_2\rangle^2}{(pp_1p_2)}}\right).
\label{ins-two}
\eea
Here we used the function $\beta_\nu(y)=\int_y^{+\infty} \de u\, u^{-\nu} e^{-\pi u}$, so that for $x\in\IR$
\be
\label{betahaf}
\beta_{\half}(x^2) = \Erfc(\sqrt{\pi} |x|),
\qquad
\beta_{\frac{3}{2}}(x^2)=2|x|^{-1}e^{-\pi x^2}-2\pi \beta_{\half}(x^2) .
\ee
In Appendix \ref{ap-indef} we provide a simple proof of the modular invariance
of $\widehat{\Psi}$ based on Vign\'eras' theorem \cite{Vigneras:1977}.
For the proof, it is important that the insertions $\Pi^{(\pm)}_{\gamma_1\gamma_2}$ cancel
the discontinuities of the sign factor $\Dela_{\gamma_1\gamma_2}$ in \eqref{defPsi}
on the loci $\cI_{\gamma_1,\gamma_2}=0$ or $\langle\gamma_1,\gamma_2\rangle=0$
in the $2b_2$-dimensional space spanned by the vectors $(\bfk_1,\bfk_2)$,
so that the summand in the completed theta series $\widehat \Psi$  is  a smooth function.
It is also important to remark that both functions \eqref{defPsih} are exponentially suppressed
as $\tau_2\to+\infty$, and that $\Pi^{(-)}_{\gamma_1\gamma_2}$ is independent of the \kahler moduli,
whereas  $\Pi^{(+)}_{\gamma_1\gamma_2}$ does depend on $t^a$ through
$\cI_{\gamma_1,\gamma_2}$ defined in \eqref{defIgg}.

In \cite{Gaiotto:2006wm,deBoer:2006vg,Denef:2007vg,Manschot:2009ia,Alexandrov:2012au}, it was
argued that the first term \eqref{defZ1} transforms as a modular form of weight
$(-\tfrac32,\tfrac12)$. As a consequence, the generating function $h_{\bfp,\bfmu}$ had
to transform as a vector-valued modular form of weight
$(-\frac{b_2}{2}-1,0)$ and multiplier system $M(\trans)=M_{\cZ}\times M_\theta^{-1}$,
where $M_\cZ=e^{2\pi\I \eps(g)p^a c_{2,a}}$ and $\eps(g)$ is the multiplier system of the Dedekind eta function.
This proposal has been confirmed in examples where the effective divisor $\cD$ wrapped
by the D3-brane is irreducible \cite{Gaiotto:2006wm,Gaiotto:2007cd},
but its validity for a general non-primitive or reducible divisor remained to be assessed.

In fact, the example of $N$ D4-branes in non-compact Calabi-Yau manifolds
(or equivalently topologically twisted $\cN=4$ $U(N)$ Yang-Mills theory \cite{Vafa:1994tf})
indicates that $h_{\bfp,\bfmu}$ is unlikely to be modular in general. In the context of
$\cN=4$ Yang-Mills, examples
are known with $\bfp$ reducible (more precisely, with the gauge group
$U(2)$ \cite{Vafa:1994tf, Minahan:1998vr, Manschot:2011dj}),
where $h_{\bfp,\bfmu}$ is not modular, but becomes so after adding to it
a suitable non-holomorphic function $R_{\bfp,\bfmu}$. In other words,
\be
\label{hath}
\whh_{\bfp,\bfmu}(\tau)=h_{\bfp,\bfmu}(\tau)-\frac{1}{2}\,R_{\bfp,\bfmu}(\tau)
\ee
is a vector-valued modular form at the cost of being non-holomorphic, while
$h_{\bfp,\bfmu}$ is a vector-valued (mixed) mock modular form  \cite{Zwegers-thesis, MR2605321}. Given
that additional non-holomorphic terms were also required to turn the double
theta series \eqref{defPsi} into a modular form $\widehat\Psi$,
we expect that for a general divisor,  the holomorphic generating function of
MSW invariants $h_{\bfp,\bfmu}$ will only become modular after the addition of a suitable
non-holomorphic function.

Assuming then that $R_{\bfp,\bfmu}$ exists such that \eqref{hath} is a vector-valued modular form,
the modular completion of the BPS partition function \eqref{defZp} becomes
\be
\widehat{\mathcal{Z}}_{\bfp}(\tau,{\bfz}, \bfc) = \sum_{n\geq 1} \widehat{\mathcal{Z}}^{(n)}_{\bfp}(\tau,{\bfz}, \bfc),
\label{defZth}
\ee
where
\be
\begin{split}
\widehat{\mathcal{Z}}^{(1)}_\bfp = &\,
\sum_{\bfmu\in\Lambda^\star/\Lambda}
\whh_{\bfp,\bfmu}\,\theta_{\bfp,\bfmu},
\\
\widehat{\mathcal{Z}}^{(2)}_\bfp = &\,
\frac{1}{2} \sum_{ \bfp_1+\bfp_2=\bfp}
\sum_{\bfmu_i\in \Lambda^\star/\Lambda_i}
\whh_{\bfp_1,\bfmu_1}\,\whh_{\bfp_2,\bfmu_2}\, \widehat \Psi_{\bfp_1,\bfp_2,\bfmu_1,\bfmu_2}.
\end{split}
\ee
Since the modular anomaly of $h_{\bfp,\bfmu}$ is expected to arise when
the divisor $\cD$ can split into several components, we expect that $R_{\bfp,\bfmu}$
should be  controlled by the product of the corresponding MSW invariants.
At this point, however, the function $R_{\bfp,\bfmu}$
remains still undetermined. We shall now fix it
by comparing the above construction with the analysis of the D3-instantons corrections to the hypermultiplet metric.

\subsection{Comparison with the contact potential}
\label{subsec-compare}

In our study of instanton effects on the hypermultiplet moduli space $\cM_H$ in the twistor formalism,
we shall find in section \ref{sec-contact} that D3-instanton contributions to the contact potential in
the two-instanton approximation can be expressed in terms of the following function:
\be
\label{defZt}
\sum_{\bfmu\in\Lambda^\star/\Lambda}
h_{\bfp,\bfmu}\,\theta_{\bfp,\bfmu}
+\frac{1}{2} \sum_{\bfp_1+\bfp_2=\bfp}
\sum_{\bfmu_i\in \Lambda^\star/\Lambda_i}
h_{\bfp_1,\bfmu_1}\,h_{\bfp_2,\bfmu_2} \(\Psi_{\bfp_1,\bfp_2,\bfmu_1,\bfmu_2}
+\Psi^{(+)}_{\bfp_1,\bfp_2,\bfmu_1,\bfmu_2}\).
\ee
In order for the metric on $\cM_H$ to carry an isometric action of the modular group, it is necessary
that \eqref{defZt} be a modular form of weight $(-\tfrac32,\tfrac12)$.

On the other hand, the completed BPS partition function \eqref{defZth} differs from
 \eqref{defZt} in two ways:
 the modular forms $\whh$ are replaced by their non-completed version $h$,
and in the second term the contribution of $\Psi^{(-)}$ is missing.
Remarkably, these two differences  cancel amongst each other provided
\be
\label{hhtoRth}
\sum_{\bfmu\in\Lambda^\star/\Lambda}
R_{\bfp,\bfmu}\,\theta_{\bfp,\bfmu}
= \sum_{\bfp_1+\bfp_2=\bfp}
\sum_{\bfmu_i\in \Lambda^\star/\Lambda_i}
h_{\bfp_1,\bfmu_1}\,h_{\bfp_2,\bfmu_2} \,\Psi^{(-)}_{\bfp_1,\bfp_2,\bfmu_1,\bfmu_2} .
\ee
In more detail, this condition ensures that the complementary terms,
appearing due to the completion of $h$ to $\whh$ in $\widehat\cZ_{\bfp}^{(1)}$,
cancel a part of the additional terms in $\widehat\cZ_{\bfp}^{(2)}$,
while the remaining discrepancy due to the difference between $h$ and $\whh$ in $\widehat\cZ_{\bfp}^{(2)}$
is of higher order in the expansion in MSW invariants. In Appendix \ref{ap-R}, we
show that the condition \eqref{hhtoRth} is solved by choosing $R_{\bfp,\bfmu}$ as in \eqref{defRmu},
where the functions $\Sg$, $ {\rm Q}_{\bfp_1,\bfp_2}$ and the variables $\tilde\bfmu$, $\bfnu_i$ are defined in
\eqref{defSgg}, \eqref{defQR}, \eqref{deftildemu} and \eqref{defnui}, respectively.
This shows that in order for the contact potential to have the right modular property,
the generating function $h_{\bfp,\bfmu}$ of MSW invariants
must have an anomalous modular transformation. Its modular completion is provided by the non-holomorphic function \eqref{defRmu},
constructed out of the generating functions $h_{\bfp_1,\bfmu_1}$ and $h_{\bfp_2,\bfmu_2}$ of MSW invariants associated to
all possible decompositions $\bfp = \bfp_1 + \bfp_2$.
In the next subsection we demonstrate that $h_{\bfp,\bfmu}$ is actually a vector-valued mixed mock modular form.

It is worth stressing that the result above is valid if $\cD$ can be written as a sum $\cD=\cD_1+\cD_2$ for {\it at most}
two effective divisors $\cD_1$ and $\cD_2$. In particular, $h_{\bfp_i,\bfmu_i}$ are modular forms,
since $\bfp_i$ cannot be further decomposed as a sum of effective charges $\bfp_i=\bfp_{i,1}+\bfp_{i,2}$.
If $\bfp$ can be written as a sum of more than two effective $\bfp_i$, then \eqref{defRmu} will involve further
corrections of higher order in MSW invariants.
It is reassuring to note that \eqref{defRmu} is consistent with explicit expressions
which are available for various non-compact Calabi-Yau's, given by
canonical bundles over a rational surface $S$.
For instance, setting $\bfp_1=\bfp_2$ in \eqref{defRmu}, it reproduces the
result  of \cite[Eq. (4.30)]{Vafa:1994tf} and \cite[Section
3.2]{Manschot:2011dj} for $\bft=-K_S$, where $K_S$ is the canonical
class of the surface $S$.

\subsection{Mock modularity and the MSW elliptic genus}
\label{subsec-modularity}

Having deduced the modular completion of the generating function of MSW invariants $h_{\bfp,\bfmu}$, which
appears as a building block of the contact potential, we shall now compare its properties with mock modular forms and
consider its implications for the elliptic genus of the MSW conformal field theory.

First we recall a few relevant aspects of mock modular forms
\cite{MR2605321, Dabholkar:2012nd}. Let $g(\tau)$ be a holomorphic modular form of weight $2-k$. The ``shadow map'' maps
$g$ to the non-holomorphic function $g^*$ defined by
\be
g^*(\tau)=(\I/2)^{k-1}\int_{-\bar \tau}^\infty (z+\tau)^{-k}\,
\overline{g(-\bar z)}\,dz.
\ee
A mock modular form of weight $k$ and with shadow $g$, is a
holomorphic function $h(\tau)$ such that its non-holomorphic completion
\be
\label{mock}
\widehat h=h+g^*
\ee
transforms as a modular form of weight $k$. Acting with the shadow
operator $\tau_2^2\partial_{\btau}$ on $\widehat h$ gives
\be
\tau_2^2\partial_{\btau} \widehat h=\tau_2^{2-k}\,\overline g\, ,
\ee
from which the shadow $g$ is easily obtained by multiplication with
$\tau_2^{k-2}$ and complex conjugation. Note that the r.h.s. transforms as a modular form of weight
$k-2$.

More generally, a mixed mock modular form of weight $k$ \cite{Dabholkar:2012nd} is a holomorphic function $h(\tau)$ such that
there exists (half) integer numbers $r_j$ and modular
forms $f_j$ and $g_j$, respectively with weights $k+r_j$ and $2+r_j$,
such that the completion
\be
\label{mixedmock0}
\widehat h=h+\sum_j\, f_j\, g_j^*
\ee
transforms as a modular form of weight $k$. Acting with the shadow operator on $\widehat h$ one obtains
\be
\label{mixedmock}
\tau_2^2\partial_{\btau} \widehat h
=\sum_j \tau_2^{2+r_j} f_j\,\overline g_j.
\ee

Let us now return to the function $h_{\bfp,\bfmu}$ and its completion $\widehat h_{\bfp,\bfmu}$ (\ref{hath}).
Applying the shadow operator $\tau_2^2 \partial_{\btau}$ to $\widehat h_{\bfp,\bfmu}$, one finds
\be
\begin{split}
\tau_2^2\p_{\btau}\,\widehat{h}_{\bfp,\bfmu} (\tau)
=&\,\frac{\sqrt{2\tau_2}}{32\pi \I} \sum_{\bfp_1+\bfp_2=\bfp}
\sum_{\bfmu_i\in\Lambda^\star/\Lambda_i}
h_{\bfp_1,\bfmu_1}(\tau)\,h_{\bfp_2,\bfmu_2}(\tau)\, \sqrt{(pp_1p_2)}
\\
&\,\times \sum_{\bfrho\in (\Lambda_1-\tilde \mu) \cap (\Lambda_2+\tilde \mu)}
(-1)^{\Sgg} e^{-2\pi\tau_2\frac{\(\Sgg\)^2}{(pp_1p_2)}+\pi\I \tau {\rm Q}_{\bfp_1,\bfp_2}\!\!\left(\bfnu_1,\bfnu_2\right)}\ ,
\end{split}
\label{shadh}
\ee
where the various symbols are defined in  Appendix \ref{ap-R}.
To see that the modular weights match on the two sides of the equation,
we observe  from (\ref{defQR}) and (\ref{defnui}) that the sum over $\bfrho$
runs over a lattice with signature $(b_2-1,1)$, therefore the second line of
(\ref{shadh}) is a theta series of weight $\frac{1}{2}(b_2-1,1)$.
Combining this with the weights of $\sqrt{\tau_2}$ and $h_{\bfp_i,\bfmu_i}$, the total weight of the
right hand side evaluates to $-(\frac{1}{2} b_2+3,0)$, consistently with the left-hand side.

Furthermore, the theta series on the second line of
(\ref{shadh}) can be expressed as a sum of holomorphic
theta series of weight $\frac{1}{2}(b_2-1)$ times anti-holomorphic
theta series of weight $\frac{1}{2}$, such that (\ref{shadh}) can be
brought to the form (\ref{mixedmock0}). This shows that $h_{\bfp,\bfmu}$ is a
(vector-valued) mixed mock modular form with $r_j=-\frac{3}{2}$ for all $j$.

To get more insight into this anomaly, we now consider
the completed elliptic genus, defined by the first term in the completed BPS partition function \eqref{defZth},
\be
\widehat \chi_{\bfp}(\tau,\bfz,\bfc)\equiv\mathcal{\widehat Z}^{(1)}_{\bfp}
=\sum_{\bfmu\in\Lambda^*/ \Lambda} \widehat h_{\bfp,\bfmu}(\tau)\,\theta_{\bfp,\bfmu}(\tau,\bft,\bfb,\bfc).
\label{compl-ellg}
\ee
The analogue of the shadow operator for $\widehat \chi_{\bfp}$
is the heat-type operator
\be
\overline{\mathcal{D}}=\tau_2^2\left( \partial_{\bar \tau}
-\frac{\I}{4\pi}\, (\partial_{c_+}+\I\pi b_+)^2 \right),
\ee
where + indicates the projection along $\bft$. This differential
operator annihilates $\theta_{\bfp,\bfmu}$ in \eqref{compl-ellg},  and therefore vanishes
 on $\widehat \chi_{\bfp}$ unless
$\widehat h_{\bfp,\bfmu}$ is non-holomorphic or, equivalently, $\bfp$
is reducible.

We shall now show that in the special case where $\bfp=2\bfp_0$ with $\bfp_0$
irreducible, the mock modularity of $h_{\bfp,\bfmu}$ is such that the ``holomorphic anomaly''
$\overline{\mathcal{D}}\,\widehat\chi_{2\bfp_0}$ is proportional to $(\chi_{\bfp_0})^2$.
Note that since we restrict to $\bfp$ which can be written as a sum
of at most two effective divisors, this implies that $\bfp_1=\bfp_2=\bfp_0$. We furthermore specialize the (real)
K\"ahler modulus $\bft$ to the large volume attractor point $\lim_{\lambda\to \infty} \lambda\,\bfp_0$ (\ref{lvolatt}),
which is the attractor point of $\bft$ at the horizon geometry AdS$_3\times S^2\times \CYm$ of the M5-brane \cite{deBoer:2008fk}.
In $\widehat\chi_{\bfp}$ the magnitude of $\bft$ is actually irrelevant, and we could just as well set $\bft=\bfp_0$.
The scale invariance also implies that the attractor points for the MSW field theories corresponding to
magnetic charges $\bfp_0$ and $2\bfp_0$ are equal.

Using these specializations in (\ref{hhtoRth}) and the expressions
(\ref{defPsih}) and (\ref{ins-two}), we find that the completion of
$\widehat \chi_{2\bfp_0}$ is obtained by adding to $\chi_{2\bfp_0}$ the following term
\be
\begin{split}
&\,\frac{1}{8\pi}\sum_{\bfmu_i\in \Lambda^*/\Lambda_i} h_{\bfp_0,\bfmu_1}(\tau)\,h_{\bfp_0,\bfmu_2}(\tau)
\sum_{\bfk_i\in \Lambda_i+\bfmu_i+\frac{1}{2}\bfp_0}
(-1)^{\bfp_0\cdot(\bfk_1+\bfk_2)+p_0^3}
|\bfp_0\cdot(\bfk_1-\bfk_2)|
\\
&\, \times \beta_{\frac{3}{2}}\!\left(\frac{\tau_2}{p_0^3}\(\bfp_0\cdot(\bfk_1-\bfk_2)\)^2\right)
\,e^{\frac{\pi\tau_2}{p_0^3}\(\bfp_0\cdot(\bfk_1-\bfk_2)\)^2}\cXt_{\bfp_0,\bfk_1}\cXt_{\bfp_0,\bfk_2}.
\end{split}
\ee
Computing the action of $\mathcal{\overline D}$ on this term, we obtain that the holomorphic anomaly of
$\widehat\chi_{2\bfp_0}$ is proportional to the square of $\chi_{\bfp_0}$,
\be
\overline{\mathcal{D}}\,\widehat\chi_{2\bfp_0}=
(-1)^{p_0^3}\,\frac{\sqrt{\tau_2 p_0^3}}{16\pi\I}\,
\chi_{\bfp_0}^2.
\ee
This extends the holomorphic anomaly of $\cN=4$ $U(2)$ gauge theory on (local)
surfaces \cite{Vafa:1994tf, Manschot:2011dj} to divisors in compact
threefolds of the form $2\bfp_0$ with $\bfp_0$ irreducible.
When the divisor $\bfp=\bfp_1+\bfp_2$ is primitive, and therefore
$\bfp_1\neq \bfp_2$, the shadow does not seem to take such a factorized form.
A possible explanation is that, whereas for $\bfp_1=\bfp_2$ the large
volume attractor points agree for magnetic charges $\bfp_1$, $\bfp_2$ and $\bfp_1+\bfp_2$,
this is not the case if $\bfp_1\neq \bfp_2$.

\section{D3-instantons in the twistor formalism}
\label{sec-twist}

In this section, we briefly review the twistorial construction of the D-instanton corrected metric
on the hypermultiplet space $\cM_H$, with emphasis on D3-instanton corrections
in the large volume limit. More details can be found in the reviews \cite{Alexandrov:2011va,Alexandrov:2013yva}
and in the original
works \cite{Alexandrov:2008nk,Alexandrov:2008gh,Alexandrov:2009zh,Alexandrov:2012au,Alexandrov:2014rca}.

\subsection{Hypermultiplet moduli space in type IIB string theory on a CY threefold}

The hypermultiplet moduli space $\cM_H$ is a \qk manifold of dimension $4b_2+4$,
which describes the dynamics of the ten-dimensional axio-dilaton $\tau=c^0+\I/g_s$,
the \kahler moduli $z^a=b^a+\I t^a$, the  Ramond-Ramond (RR) scalars $c^a,\cla,\cl0$, corresponding to periods of the RR
2-form, 4-form and 6-form on a basis of $H^{\rm even}(\CYm,\IZ)$, and finally,
the NS axion $\psi$, dual to the Kalb-Ramond two-form $B$ in four dimensions. At tree-level,
the metric on $\cM_H$ is obtained from the moduli space $\cM_{\cS\cK}$ of
complexified K\"ahler deformations via the
$c$-map construction \cite{Cecotti:1988qn,Ferrara:1989ik}.
The special \kahler manifold
$\cM_{\cS\cK}$ is characterized by the holomorphic prepotential $F(X)$
where $X^\Lambda$ are homogeneous complex coordinates on $\cM_{\cS\cK}$
such that $X^\Lambda/X^0=z^\Lambda$ (with $z^0=1$).
Classically, the prepotential is determined by the intersection numbers
$
\label{lve}
\Fcl(X)=-\kappa_{abc}\frac{X^a X^b X^c}{6 X^0},
$
whereas quantum mechanically it is affected by $\alpha'$-corrections,
which are however suppressed in the large volume limit.

As mentioned in the introduction, beyond the tree-level the metric on $\cM_H$ receives quantum $g_s$-corrections.
At the perturbative level there is only a one-loop correction, proportional to the Euler characteristics $\chi_{\CYm}$.
The corresponding metric is a one-parameter deformation of the $c$-map found explicitly in a series of works
\cite{Antoniadis:1997eg,Gunther:1998sc,Antoniadis:2003sw,Robles-Llana:2006ez,Alexandrov:2007ec}.
At the non-perturbative level, there are corrections from D-branes
wrapped on complex cycles in $\CYm$ (described by coherent sheaves on $\CYm$),
and from NS5-branes wrapped on $\CYm$, which we ignore in this paper.

Before recalling how D-brane instantons affect the metric,  a few words are in order about the symmetries of $\cM_H$.
In the classical, large volume limit, the metric is invariant
under the semi-direct product of $SL(2,\IR)$ times the graded nilpotent algebra
$N=N^{(1)}\oplus N^{(2)}\oplus N^{(3)}$, where the generators in $N^{(1)}, N^{(2)}, N^{(3)}$
transform as $b_2$ doublets, $b_2$ singlets and one doublet under $SL(2,\IR)$, respectively. Quantum corrections
break this continuous symmetry, but are expected to preserve an isometric action of the discrete subgroup $SL(2,\IZ)\ltimes N(\IZ)$,
where $SL(2,\IZ)$ descends from S-duality group of type IIB supergravity,
while the nilpotent factor corresponds to monodromies around the large volume point
and large gauge transformations of the RR and Kalb-Ramond fields.
Under an element $g={\scriptsize \begin{pmatrix} a & b \\ c & d \end{pmatrix}} \in SL(2,\IZ)$,
the type IIB fields transforms as in \eqref{SL2Z},
supplemented by the following action on $\cla, \cl0, \psi$,
\be\label{SL2Zc}
\cla\mapsto \cla - c_{2,a} \varepsilon(g)\ ,
\qquad
\begin{pmatrix} \cl0 \\ \psi \end{pmatrix} \mapsto
\begin{pmatrix} d & -c \\ -b & a  \end{pmatrix}
\begin{pmatrix} \cl0 \\ \psi \end{pmatrix}\ ,
\ee
where $\varepsilon(g)$ is the logarithm of the multiplier system of the Dedekind eta function
\cite{Alexandrov:2010ca,Alexandrov:2012au}. In the absence of D5 and NS5-brane instantons,
the metric admits two additional continuous
isometries, acting by shifts of $\tc_0$ and $\psi$.

\subsection{Twistorial construction of D-instantons}

Quantum corrections to $\cM_H$ are most easily described using the language of twistors.
The twistor space $\cZ$ of $\cM_H$ is a $\CP$-bundle over $\cM_H$ endowed with a complex contact structure.
This contact structure is represented by a (twisted) holomorphic one-form $\cX$,
which locally can always be expressed in terms of complex Darboux coordinates as
\be
\cX^{[i]}=  \de\ai{i}+ \txii{i}_\Lambda \de \xii{i}^\Lambda\, ,
\label{contform}
\ee
where the index $\scriptsize [i]$ labels the
patches $\cU_i$ of an open covering of $\CP$. The global  contact structure on $\cZ$
(hence, the metric on $\cM$)  is then encoded in contact transformations between
Darboux coordinate systems on the overlaps $\cU_i \cap \cU_j$.
It is convenient to parametrize these transformations by holomorphic functions $\Hij{ij}(\xi^\Lambda,\txi_\Lambda,\alpha)$,
known as contact hamiltonians\footnote{The contact hamiltonians coincide
with the generating functions introduced in \cite{Alexandrov:2008nk}
in the special case where $\Hij{ij}$ is independent of $\txi_\Lambda$ and $\alpha$.} \cite{Alexandrov:2014mfa},
which generate the contact transformations by exponentiating
the action of the vector field
\be
\begin{split}
X_H = \left( -\p_{\txi_\Lambda} H+\xi^\Lambda\p_\alpha H \right) \partial_{\xi^\Lambda}
+ \p_{\xi^\Lambda} H\, \partial_{\txi_\Lambda} +
\left( H-\xi^\Lambda\p_{\xi^\Lambda} H \right) \partial_{\alpha}.
\end{split}
\label{contbr}
\ee
Thus, a set of such holomorphic functions associated to a covering of $\CP$
(satisfying obvious consistency conditions on triple overlaps)
uniquely defines a \qk manifold.

To extract the metric from these data, the first step is to express the Darboux coordinates
in terms of coordinates on $\cM_H$ and the stereographic coordinate $t$ on $\CP$.
They are fixed by regularity properties and the gluing conditions
\be
(\xii{j}^\Lambda, \txii{j}_\Lambda ,\ai{j}) = e^{X_{\Hij{ij}}} \cdot (\xii{i}^\Lambda, \txii{i}_\Lambda , \ai{i}),
\ee
which typically can be rewritten as a system of integral equations.
Once the Darboux coordinates are found, it is sufficient to plug them into the contact one-form \eqref{contform},
expand around any point $t\in \CP$ and read
off the components of the $SU(2)$ part of the Levi-Civita connection $\vec p$.  E.g. around
the point $t=0$, the expansion reads
\be
\cX^{[i]}= -4\I\, e^{\Phi^{[i]}} \left( \frac{\de t}{t} + \frac{p_+}{t} - \I \, p_3 + p_- t \right).
\label{def-oneform}
\ee
The scale factor $e^{\Phi^{[i]}}$ is known as the contact potential \cite{Alexandrov:2008nk}.
In the case when $\cM_H$ has a continuous isometry and
the Darboux coordinates are chosen such that this isometry lifts to the vector field $\partial_{\alpha}$,
$\Phi$ is globally well-defined (i.e. independent of the patch index $\scriptsize [i]$) and is independent of the fiber coordinate $t$.
Thus, it becomes a function on $\cM_H$ which, in fact, coincides with the norm of the moment map associated to
the isometry \cite{Alexandrov:2011ac}.

In this formalism the D-instanton corrected hypermultiplet moduli space, with NS5-brane instantons being ignored,
was constructed in \cite{Alexandrov:2008gh,Alexandrov:2009zh}. We omit the details of this construction and
present only those elements which are relevant for the analysis of the contact potential.
\begin{itemize}
\item
The contact hamiltonians enforcing D-instanton corrections to the metric are given by
\be
H_\gamma(\xi,\txi)= \frac{\bar\Omega(\gamma)}{(2\pi)^2}\,\sigma_\gamma\cX_\gamma\, ,
\label{prepHnew}
\ee
where $\cX_\gamma = \expe{p^\Lambda \txi_\Lambda-q_\Lambda \xi^\Lambda}$,
$\bar\Omega(\gamma)$ are rational DT invariants \eqref{defntilde}, and $\sigma_\gamma$
is a quadratic refinement of the intersection pairing on $H^{\rm even}(\CYm)$, a sign factor which we fix below.
They generate contact transformations connecting Darboux coordinates
on the two sides of the BPS rays $\ell_\gamma$ on $\CP$ extending from $t=0$ to $t=\infty$,
along the direction fixed by the central charge
\be
\ell_\gamma = \{ \varpi\in \CP\ :\ \ Z_\gamma/\varpi\in \I \IR^-\}  .
\ee

\item
The Darboux coordinates are obtained by solving the following integral equations
\be
\cX_\gamma(t) = \cXsf_\gamma(t)\, \expe{
\frac{1}{8\pi^2}
\sum_{\gamma'} \sigma_{\gamma'}\bar\Omega(\gamma')\, \langle \gamma ,\gamma'\rangle
\int_{\ell_{\gamma'} }\frac{\de t'}{t'}\, \frac{t+t'}{t-t'}\,
\cX_{\gamma'}(t')},
\label{eqTBA}
\ee
where
\be
\cXsf_\gamma(t)=\expe{\frac{\tau_2}{2}\(\bZ_\gamma(\ub)\,t-\frac{Z_\gamma(u)}{t}\)
+p^\Lambda \tzeta_\Lambda- q_\Lambda \zeta^\Lambda}
\label{defXsf}
\ee
are the Fourier modes of the tree-level (or `semi-flat')
Darboux coordinates valid in the absence of D-instantons.\footnote{The argument $u^a$ of
the central charge $Z_\gamma$ can be understood as the complex structure moduli in the mirror threefold and
will be related to the \kahler moduli $z^a$ of the threefold $\CY$ by the mirror map.}
In the weak coupling limit, these equations can be solved iteratively, leading to a (formal) multi-instanton series.
This gives $\xi^\Lambda$ and $\txi_\Lambda$
in each angular sector, which can then be used to compute
the  Darboux coordinate $\alpha$, whose explicit expression can be found in \cite{Alexandrov:2009zh}
and will not be needed in this paper.
These equations provide Darboux coordinates
as functions of the fiber coordinate $t$ and variables
$(\tau_2, u^a, \zeta^\Lambda, \tzeta_\Lambda,\sigma)$
which play the role of coordinates on $\cM_H$. They are adapted to the symmetries of the type IIA formulation
and therefore can be considered as natural coordinates on the moduli space of type IIA  string theory compactified
on the mirror Calabi-Yau threefold. Their relation to the type IIB fields will be explained in the next subsection.

\item
Given the Darboux coordinates $\cX_\gamma$, the contact potential $e^\Phi$ is obtained from the Penrose-type integrals
\be
e^{\Phi} = \frac{\I\tau_2^2}{16}\(\ub^\Lambda F_\Lambda- u^\Lambda \bF_\Lambda\)-\frac{\chi_{\CYm}}{192\pi}
+\frac{\I\tau_2}{64\pi^2}\,\sum_\gamma \sigma_\gamma\Omega(\gamma)
\int_{\ell_\gamma} \frac{\text{d}t}{t} \( t^{-1} Z_\gamma(u)-t\bZ_\gamma(\ub)\) \cX_\gamma.
\label{phiinstmany}
\ee

\end{itemize}

\subsection{S-duality and mirror map}

The D3-instanton corrected metric is obtained from the construction above by
assuming that the only non-vanishing DT invariants $\Omega(p^\Lambda,q_\Lambda;z^a)$
are those where the  D5-brane charge $p^0$ vanishes. While we expect that
this metric  should carry an isometric action of $SL(2,\IZ)$, this
symmetry is far from being manifest. Indeed, the construction above
is adapted to symplectic invariance, which is manifest in type IIA formulation, rather than to S-duality,
which is explicit on the type IIB side.
In particular, the Darboux coordinates are defined by \eqref{eqTBA}
in terms of type IIA variables. In order to understand their behavior under S-duality,
 they should be rewritten instead in terms of type IIB variables, which we {\it define} by their transformation
properties \eqref{SL2Z}, \eqref{SL2Zc}. We refer to the change of coordinates from IIA to IIB variables as the {\it mirror map}.

In the classical approximation (i.e. tree-level, large volume limit), the mirror map was found in \cite{Bohm:1999uk}
and is given by
\be
\label{symptobd}
\begin{split}
u^a & =b^a+\I t^a\, ,
\qquad
\zeta^0=\tau_1\, ,
\\
\tzeta_a &=  \cla+ \frac{1}{2}\, \kappa_{abc} \,b^b (c^c - \tau_1 b^c)\, ,
\qquad
\tzeta_0 = \cl0-\frac{1}{6}\, \kappa_{abc} \,b^a b^b (c^c-\tau_1 b^c)\, ,
\\
\sigma &= -(2\psi+  \tau_1 \cl0) + \cla (c^a - \tau_1 b^a)
-\frac{1}{6}\,\kappa_{abc} \, b^a c^b (c^c - \tau_1 b^c)\, .
\end{split}
\ee
One can check that, if one substitutes these expressions into the classical Darboux coordinates,
obtained by dropping all integrals and retaining only the classical part the holomorphic prepotential $F(X)$,
and supplement the $SL(2,\IZ)$ transformations of the type IIB fields by
the following fractional transformation of the fiber coordinate
\be
\varpi \mapsto  \frac{c \tau_2 + \varpi  (c \tau_1 + d) +
\varpi |c \tau + d| }{(c \tau_1 + d) + |c \tau + d| - \varpi c \tau_2}\, ,
\label{transz}
\ee
the resulting Darboux coordinates transform holomorphically as in \cite[Eq.(2.20)]{Alexandrov:2012au}.
Moreover, it is straightforward to check that this transformation preserves the contact structure
since it rescales the contact one-form by a holomorphic factor,
\be
\cX\mapsto\frac{\cX}{c\xi^0+d}.
\label{tr-contact}
\ee
This demonstrates that $SL(2,\IZ)$ acts isometrically on $\cM_H$ in the classical approximation.

To go beyond this approximation, one must ensure that, even after inclusion of quantum corrections
into the Darboux coordinates, $SL(2,\IZ)$ still acts on them by a holomorphic contact transformation.
The main complication comes from the fact that the mirror map itself gets corrected.
Thus, the key problem is to find corrections to \eqref{symptobd} such that the resulting Darboux coordinates
transform holomorphically and the contact one-form satisfies \eqref{tr-contact}.

For the pure D1-D(-1)-instantons this problem was solved in \cite{Alexandrov:2009qq}.
Furthermore, it was shown that after a local contact transformation,
the instanton corrected Darboux coordinates transform exactly as the classical ones,
and the description of the twistor space in terms of a covering of $\CP$ and
contact hamiltonians takes a manifestly modular invariant form.
Then in \cite{Alexandrov:2012bu} it was understood how to derive the mirror map for generic
QK manifolds obtained by a deformation of the c-map and preserving two-continuous isometries,
of which $\cM_H$ corrected by D3-instantons is a particular case.\footnote{In this class of geometries
the action of $SL(2, \IZ)$ on the fiber coordinate \eqref{transz} remains uncorrected, unlike
in the case without any continuous isometries considered in \cite{Alexandrov:2013mha}.}
The idea is that the mirror map is induced by converting the kernel
$\frac{\de \varpi'}{ \varpi'} \,\frac{\varpi'+\varpi}{\varpi'-\varpi}$,
appearing in the integral equations for Darboux coordinates and transforming non-trivially under S-duality,
into a modular invariant kernel.
In \cite{Alexandrov:2012au} these results were applied to D3-instantons in the one-instanton, large volume approximation.
We summarize them in the next subsection. But before that we make two comments.

First, it is convenient to redefine the coordinate $t$ by
a Cayley transformation:
\be
\label{Cayley}
z=\frac{t+\I}{t-\I}\, .
\ee
The transformation \eqref{transz}, lifting the $SL(2,\IZ)$ action on $\cM_H$ to a holomorphic
contact transformation in twistor space, then takes the simpler form
\be
\label{ztrans}
z\mapsto \frac{c\bar\tau+d}{|c\tau+d|}\, z\,  .
\ee
In particular, the two points $t=\mp\I$ on $\CP$, which stay invariant under the $SL(2,\IZ)$ action,
are mapped to $z=0$ and $z=\infty$, which  makes it easier to do a Fourier expansion around them.
Secondly, the fact that $\frac{\de \varpi}{\varpi}$ transforms into $\frac{|c\tau+d|}{c\xi^0+d}\,\frac{\de \varpi}{\varpi}$
under \eqref{transz}, along with \eqref{def-oneform} and \eqref{tr-contact}, shows that
the contact potential $e^\Phi$ must  transform as a modular form of weight $(-\tfrac12, -\tfrac12)$ \cite{Alexandrov:2008gh},
\be
\label{SL2phi}
e^\Phi \mapsto \frac{e^\Phi}{|c\tau+d|}\, .
\ee

\subsection{D3-instantons in the one-instanton approximation: a summary}

Here we summarize the results obtained in \cite{Alexandrov:2012au} which are relevant for the evaluation of
the contact potential in the next section.
First, we recall that they are derived in the one-instanton, large volume approximation,
which means that one restricts to the first order in the expansion in the DT or MSW invariants
and takes the limit $t^a\to\infty$. In this limit, the integrals along BPS rays $\ell_\gamma$
are dominated by a saddle point at
\be
z'_\gamma\approx
-\I\frac{(q+b)_+}{\sqrt{(pt^2)}}
\ee
for $(pt^2)>0$ and $z'_{-\gamma}=1/z'_\gamma$ in the opposite case. This shows that in all integrands
we can send $z'$ either to zero or infinity keeping constant $t^a z' $ or $t^a/z'$, respectively.

Next, we should fix the quadratic refinement appearing in \eqref{prepHnew}. We choose it to be
$\sigma_\gamma=\expe{\hf\, p^a q_a}\sigma_\bfp$ where $\sigma_\bfp=\expe{\hf\, A_{ab}p^ap^b}$
and $A_{ab}$ is the matrix satisfying
\be
\label{ALS}
A_{ab} p^p - \frac12\, \kappa_{abc} p^b p^c\in \IZ \quad \text{for}\ \forall p^a\in\IZ,
\ee
and performing the symplectic rotation making the charges integer valued (see footnote \ref{foot-charges}).
It is easy to check that such quadratic refinement satisfies the defining relation
\be
\sigma_{\gamma_1}\sigma_{\gamma_2}=(-1)^{\langle\gamma_1,\gamma_2\rangle}\sigma_{\gamma_1+\gamma_2}.
\ee

With these definitions one has the following results:
\begin{itemize}
\item {\it Quantum mirror map.}
The following expression
\be
\begin{split}
u^a=&\,b^a+\I t^a -\frac{\I}{8\pi^2\tau_2}{\sum_{\gamma\in\Gamma_+}}\sigma_\gamma \bar\Omega(\gamma)p^a\[
\int_{\ell_{\gammap}}\de z\,(1-z)\,\cX_{\gammap}
+\int_{\ell_{\gammam}}\frac{\de z}{z^3} (1-z)\,\cX_{\gammam}\]
\end{split}
\label{inst-mp}
\ee
replaces the simple relation \eqref{symptobd} between the complex structure moduli and the type IIB fields.
The other relations of the classical mirror map also get corrections due to D3-instantons,
but are not needed for the purposes of this paper.

\item {\it Darboux coordinates.}
If one defines the instanton expansion of the Fourier modes $\cX_\gamma$ as
\be
\cX_\gamma=\cXcl_\gamma\(1+\cXq_\gamma+\cdots\),
\label{expcX}
\ee
then for $\gamma\in\Gamma_+$ one finds\footnote{Although \eqref{cXone} did not appear explicitly in \cite{Alexandrov:2012au},
it can be easily obtained from Eqs. (4.5) and (4.7) of that paper.}
\bea
\vphantom{A\over A_A}
\cXcl_\gamma &=&e^{-2\pi S^{\rm cl}_\bfp}\,
\expe{ - \hat q_0\tau+\I Q_{\gammap}(z)}
\cXt_{\bfp,\bfq},
\label{cXzero}\\
\cXq_{\gammap}&=&
\frac{1}{2\pi}\sum_{\gamma'\in\Gamma_+}\sigma_{\gamma'}\Omega(\gamma')\int_{\ell_{\gamma'}}\de z'
\((tpp')-\frac{\I\langle\gamma,\gamma'\rangle}{z'-z}\)\cXcl_{\gamma'},
\label{cXone}
\eea
where $\cXt_{\bfp,\bfq}$ was defined in \eqref{defXt} and gives rise to the usual Siegel theta series,
$S^{\rm cl}_\bfp$ is the leading part of the Euclidean D3-brane action
in the large volume limit, and  $Q_{\gamma}(z)$ is the only part of $\cXcl_\gamma$ depending on $z$,
\be
S^{\rm cl}_\bfp = \frac{\tau_2}{2}\,(pt^2) - \I  p^a \tc_a ,
\qquad
Q_{\gammap}(z)=\tau_2 (pt^2)
\left(z + \I\, \frac{ \qbp}{\sqrt{(pt^2)}}\right)^2.
\label{clactinst}
\ee
Note that $\cXcl_\gamma$ is the part of $\cXsf_\gamma$ \eqref{defXsf} obtained by using the classical mirror map \eqref{symptobd},
whereas $\cXq_{\gamma}$ has two contributions: one from the integral term in the equation \eqref{eqTBA}
and another from quantum corrections to the mirror map.
For the opposite charge, the results can be obtained via the complex conjugation and the antipodal map,
$\cX_{-\gamma}(z)=\overline{\cX_\gamma(-\bz^{-1})}$.

\item {\it Contact potential.}
The D3 one-instanton contribution to the contact potential takes the simple form
\be
\delta^{(1)}e^\Phi= \frac{\tau_2}{2}\Re\sum_{\bfp}\,\cD_{-\frac{3}{2}}\cFq_\bfp,
\label{del1Phi}
\ee
where
\be
\cD_{\wh} = \frac{1}{2\pi \I}\(\partial_\tau+\frac{\wh}{2\I\tau_2}+ \frac{\I t^a}{4\tau_2}\, \partial_{t^a}\),
\label{modcovD}
\ee
is the modular covariant derivative operator mapping
modular functions of weight $(\wh,\bwh)$ to modular functions of weight $(\wh+2,\bwh)$,
whereas the function $\cFq_\bfp$ is given by
\bea
\cFq_{\bfp}&=& \frac{1}{4\pi^2}\sum_{q_\Lambda}\sigma_\gamma \bar\Omega(\gamma)\int_{\ell_{\gammap}} \de z\, \cXcl_{\gammap}
=\frac{\sigma_{\bfp}\, e^{-\pi\tau_2 (pt^2)+2\pi\I p^a\tc_a }}{4\pi^2  \sqrt{2\tau_2(pt^2)}}\, \mathcal{Z}^{(1)}_\bfp
\label{defcFq}
\eea
and is proportional to the MSW elliptic genus \eqref{defZ1}. Since in this approximation, $\mathcal{Z}^{(1)}_\bfp$
transforms as a modular form of weight $(-\tfrac32,\tfrac12)$, the contact potential satisfies the required modular properties.

\end{itemize}

\section{Contact potential at two-instanton order}
\label{sec-contact}

Although the complete proof that the D3-instanton corrected hypermultiplet
moduli space carries an isometric action of $SL(2,\IZ)$
requires analyzing the modular transformations of the full system of Darboux coordinates,
in this paper we restrict our attention to the modular properties of the contact potential.
This provides a highly non-trivial test of S-duality and, as was explained in section \ref{subsec-modularity},
already has important implications for understanding the modular properties
of the elliptic genus and the partition function of DT invariants.
The analysis of Darboux coordinates will be presented in \cite{abmp-to-appear}.

Let us evaluate the D3-instanton contribution to the contact potential \eqref{phiinstmany} up to second order
in the expansion in DT invariants. This requires the knowledge of the mirror map for $u^a$ to the same order.
We assume that it is given by the same relation \eqref{inst-mp} as above where, however,
the Fourier modes $\cX_\gamma$ should now be substituted by their expressions \eqref{expcX} including one-instanton contributions.
In appendix \ref{ap-contact} we provide the details of the calculation.
To present the final result, we first introduce the obvious generalization of the function \eqref{defcFq},
\be
\cF=\frac{1}{4\pi^2}\sum_{\gamma\in\Gamma_+}\sigma_\gamma \bar\Omega(\gamma)\int_{\ell_{\gammap}} \de z\, \cX_{\gammap}.
\label{defcF}
\ee
Expanding it to the second order, one finds
\be
\cF=\sum_{\bfp}\cFi{1}_{\bfp}+\sum_{\bfp_1,\bfp_2}\cFi{2}_{\bfp_1\bfp_2}+\cdots\, ,
\label{expcF}
\ee
where $\cFi{1}_{\bfp}$ is given in \eqref{defcFq}, whereas the second order term reads
\be
\cFi{2}_{\bfp_1\bfp_2}=\frac{1}{8\pi^3}\sum_{q_{1,\Lambda},q_{2,\Lambda}}
\sigma_{\gamma_1}\sigma_{\gamma_2}\bar\Omega(\gamma_1)\bar\Omega(\gamma_2)
\int_{\ell_{\gammap_1}} \!\!\de z_1\,\int_{\ell_{\gammap_2}} \!\! \de z_2
\[(tp_1p_2)-\frac{\I\langle\gamma_1,\gamma_2\rangle}{z_2-z_1} \]\cXcl_{\gammap_1}(z_1)\cXcl_{\gammap_2}(z_2),
\label{defcF2}
\ee
where we used \eqref{cXone}.  The function encoding the contact potential is
obtained by {\it halving} the coefficient of the second order term in the simple twistor integral \eqref{defcF},
\be
\tcF_\bfp=\cFi{1}_{\bfp}+\hf\sum_{\bfp_1+\bfp_2=\bfp}\cFi{2}_{\bfp_1\bfp_2},
\label{deftcF}
\ee
While this prescription may seem surprising at first sight, it follows from the indistinguishability
of the constituents with charges $\bfp_1, \bfp_2$.

In terms of the function \eqref{deftcF}, the D3-instanton contribution to the contact potential has a simple representation
which generalizes \eqref{del1Phi},
\be
\delta e^\Phi =\frac{\tau_2}{2}\Re\sum_{\bfp}\,\cD_{-\frac{3}{2}}\tcF_\bfp
-\frac{1}{8}\sum_{\bfp_1,\bfp_2}(tp_1p_2)\,\cFq_{\bfp_1}\overline{\cFq_{\bfp_2}}.
\label{Phitwo-Dp}
\ee
Since $\cFq_{\bfp_1}$ transforms under S-duality with modular weight $(-\tfrac32,\tfrac12)$\footnote{The modular anomaly of $\cFq_{\bfp_1}$
discussed in section \ref{sec-BPS} is of second order in the instanton expansion and therefore can be ignored in our approximation
in the discussion of the last term in \eqref{Phitwo-Dp}.}, it is immediate to see that the last term in \eqref{Phitwo-Dp}
transforms as a modular form of weight $(-\half,-\half)$, as required for the contact potential.
Hence the same should be true for the first term. Since $\cD_{-\frac{3}{2}}$ is a modular covariant operator,
raising the weight by (2,0), one concludes that modular invariance requires that the full function $\tcF_\bfp$
must be a modular form of weight $(-\tfrac32,\tfrac12)$.

Let us rewrite this function in terms of MSW invariants and perform its theta series decomposition.
To this end, we plug in the expansion of DT invariants given by \eqref{multiMSW}, keeping only terms of second order in $\bOmMSW$,
and then substitute \eqref{cXzero}. The result is
\bea
\tcF_{\bfp}&=& \frac{ e^{-2\pi S^{\rm cl}_\bfp}}{4\pi^2}\Biggl\{
\sum_{\bfmu \in \Lambda^\star/\Lambda}\,h_{\bfp,\bfmu}
\sum_{\bfk\in\Lambda_i+\bfmu+\half \bfp} \sigma_\gamma\cY_\gamma\,\cXt_{\bfp,\bfk}
\nn\\
&&
+ \hf\sum_{ \bfp_1+\bfp_2=\bfp}\sum_{\bfmu_i\in \Lambda^\star/\Lambda_i}
h_{\bfp_1,\bfmu_1}\,h_{\bfp_2,\bfmu_2}\sum_{\bfk_i\in\Lambda_i+\bfmu_i+\half \bfp_i}\sigma_{\gamma_1}\sigma_{\gamma_2}
\biggl[\langle\gamma_1,\gamma_2 \rangle \Dela_{\gamma_1\gamma_2}\cY_\gamma
\expe{\frac{\tau}{2}\, {\rm Q}_{\bfp_1,\bfp_2}(\bfk_1,\bfk_2)}\cXt_{\bfp,\bfk}
\biggr.
\nn\\
&& \biggl.\qquad
+\frac{1}{2\pi}\,\Bigl((tp_1p_2)\,\cY_{\gamma_1}\cY_{\gamma_2}
-\I\langle\gamma_1,\gamma_2\rangle\, \dInt \Bigr)\biggr]\cXt_{\bfp_1,\bfk_1}\cXt_{\bfp_2,\bfk_2} \Biggr\},
\label{defcFqthet}
\eea
where we defined
\be
\label{defdInt}
\cY_\gamma=\int_{\ell_\gamma} \de z\, e^{-2\pi Q_{\gamma_1}(z)},
\qquad
\dInt=\int_{\ell_{\gamma_1}}\!\! \de z_1 \int_{\ell_{\gamma_2}} \!\!\de z_2\,
\frac{e^{-2\pi \(Q_{\gamma_1}(z_1)+Q_{\gamma_2}(z_2)\)}}{z_2-z_1}.
\ee
The first integral is Gaussian and is easily evaluated, whereas the second integral is more involved and
computed in appendix \ref{ap-doubleintcont}. The two results are
\be
\cY_\gamma=\frac{1}{\sqrt{2\tau_2(pt^2)}}\, ,
\qquad
\dInt=-\frac{\pi \I \,e^{2\pi\tau_2 \cI_{\gamma_1\gamma_2}^2}}{\sqrt{2\tau_2(pt^2)}}\,
\sgn(\cI_{\gamma_1\gamma_2})\,\beta_{\half}(2\tau_2\cI^2_{\gamma_1\gamma_2}).
\label{defdIntcp}
\ee
Plugging them into \eqref{defcFqthet} and taking into account that due to \eqref{ALS}
\be
\sigma_{\bfp_1}\sigma_{\bfp_2}\sigma_{\bfp_1+\bfp_2}^{-1}=\expe{A_{ab}p_1^a p_2^b}=(-1)^{(p_1^2p_2)},
\ee
one finds that $\tcF_\bfp$ coincides, up to a modular invariant factor, with the function given in \eqref{defZt}.
As was shown in section \ref{sec-BPS}, this function is indeed a modular form of weight $(-\tfrac32,\tfrac12)$
provided $h_{\bfp,\bfmu}$ is mock modular and its modular completion is given by \eqref{hath} and \eqref{defRmu}.
In this case, in the second order approximation, it can be identified with the partition function \eqref{defZth}
so that we arrive at
\be
\tcF_{\bfp}=\frac{\sigma_{\bfp}\, e^{-2\pi S^{\rm cl}_\bfp}}{4\pi^2  \sqrt{2\tau_2(pt^2)}}\,\widehat{\mathcal{Z}}_\bfp\, .
\ee
This result generalizes \eqref{defcFq} and ensures the right transformation properties of the contact potential.

\section{Discussion}
\label{sec-concl}

In this paper we studied the  invariance of the  D3-instanton corrected metric on the
hypermultiplet moduli space in type IIB  Calabi-Yau string vacua under S-duality.
We restricted ourselves to the two-instanton, large volume
approximation and concentrated on the contact potential $e^\Phi$, which is sensitive to all quantum
corrections to the metric. S-duality requires that it  must transform as a modular form of fixed weight.
We showed that in our approximation, D3-instanton contributions to $e^\Phi$ can be
expressed in terms of the modular derivative of a function $\widehat{\mathcal{Z}}_\bfp$
constructed, on the one hand, from a non-holomorphic modification $\whh_{\bfp,\bfmu}$ \eqref{hath}
of the generating function of the MSW invariants
$h_{\bfp,\bfmu}$, and on the other hand, from the double theta series $\widehat\Psi$ constructed in \cite{Manschot:2009ia}.
The modular invariance of $e^\Phi$ then requires that $\widehat{\mathcal{Z}}_\bfp$ should transform
as a modular form (of fixed weight and multiplier system), which in turn implies that $\whh_{\bfp,\bfmu}$
should transform as a vector-valued modular form.
Thus, when the divisor $\cD$ wrapped by the D3-brane is reducible, $h_{\bfp,\bfmu}$
is only mock  modular. In the case when $\cD$ can split into at most two effective divisors
$\cD_1+\cD_2$, the modular anomaly is dictated by the non-holomorphic completion $R_{\bfp,\bfmu}$ given in \eqref{defRmu}.
Clearly, it would be desirable to find  independent checks of this
conjecture. Beyond this, our work opens several avenues for future research:

\begin{itemize}
\item
The modular function $\widehat \CZ_\bfp$ \eqref{defZthintro} can be viewed as the BPS partition
function of D4-D2-D0 black holes in $\cN=2$ supergravity in $\mathbb{R}^{3,1}$.
The non-holomorphic terms which are necessary for modularity, can be understood as a consequence
of the continuum of scattering states in $\mathbb{R}^{3,1}$. In particular, the
term proportional to $\beta_{\half}\!\left(2\tau_2\cI_{\gamma_1\gamma_2}^2 \right)$ in \eqref{ins-one}
is recognized
as the contribution of the continuum to the $\IR^3$-index \cite{Alexandrov:2014wca} computed in  \cite[(3.37)]{Pioline:2015wza},
upon identifying $R=\tau_2, \cI_{\gamma_1\gamma_2}^2=|Z_{\gamma_1}|+|Z_{\gamma_2}|-|Z_{\gamma_1+\gamma_2}|=\tfrac{\vartheta^2}{2m}$, in agreement
with the observation in footnote \ref{foobind}. It would be interesting to understand
the origin of the first term in \eqref{ins-one}, which was not seen in the analysis 
of \cite{Pioline:2015wza}.

\item
As was done in \eqref{compl-ellg}, the first term $\widehat \CZ_\bfp^{(1)}$ in \eqref{defZthintro} should be interpreted
as the completed elliptic genus $\widehat\chi_{\bfp}$ of the MSW SCFT obtained by wrapping an M5-brane on $\cD$.
The holomorphic anomaly is expected to arise from a spectral asymmetry in a continuum of states,
corresponding to configurations where the M5-brane
on $\cD$ splits into two M5-branes wrapping $\cD_1$ and $\cD_2$. It would be interesting to
derive the modular anomaly of the generating function $h_{\bfp,\bfmu}$ using this interpretation,
which would require a better understanding of the M5-brane CFT \cite{Maldacena:1997de,Minasian:1999qn}.

\item Another challenging approach is to determine the DT invariants
$\Omega(\gamma;\bfz)$ for specific
Calabi-Yau manifolds using enumerative algebraic geometry. Since our results
give an explicit constraint
on DT invariants for pure codimension 2 sheaves, the knowledge
of $\Omega(\gamma;\bfz)$ will allow to verify the necessity of the non-holomorphic modification
$R_{\bfp,\bfmu}$. One could, for example, try to extend the approach of \cite{Gaiotto:2006wm} to two
D4-branes on the quintic, or to Calabi-Yau threefolds with $b_2>1$.

\item Returning to  the subject of hypermultiplet moduli spaces, the fact that the contact potential
is modular is only a necessary condition for the existence of an isometric action
of S-duality on $\cM_H$.
A complete proof requires  analyzing the Darboux coordinates and  showing that they transform
by a holomorphic
contact transformation, as was done in the one-instanton approximation in \cite{Alexandrov:2012au}.
A similar analysis at two-instanton level will be the subject of the subsequent work  \cite{abmp-to-appear}.

\item It would also be interesting to go beyond the large volume approximation and to arbitrary order in the multi-instanton expansion.
An important step in this direction would be to obtain a twistorial formulation of D3-instantons which is manifestly invariant under S-duality,
along the lines outlined in \cite{Alexandrov:2012bu}.

\item Finally, a far-reaching goal is to get a complete non-perturbative description of the exact
hypermultiplet moduli space
including, in particular, five-brane instantons.
Their modularity can be enforced by covariantizing the known results on D5-instantons under S-duality.
This was realized at a linearized level in \cite{Alexandrov:2010ca} and attempted beyond the linear approximation
in \cite{Alexandrov:2014mfa,Alexandrov:2014rca},
but it is not clear whether a simple covariantization is sufficient to get the complete picture.

\end{itemize}

\medskip

{\noindent \bf Acknowledgments:}
We would like to thank N. Gaddam, A. Klemm, S. Murthy and S. Vandoren for useful discussions.
S.A. is grateful to the CERN Theory Group and to IHES,
S.B. to the CERN Theory Group and to L2C in Montpellier, and
B.P. to the Math Department at Trinity College Dublin and to L2C in Montpellier for hospitality during a part of this project.
S.B. also thanks CEFIPRA for financial support.

\appendix

\section{Indefinite theta series}
\label{ap-indef}

In this section, we provide an alternative proof to the fact that $\widehat\Psi=\Psi+\Psi^{(+)}+\Psi^{(-)}$ transforms as
a vector valued modular form of weight $(b_2+\frac12,\frac12)$.
The original alternative proof, which can be found in Proposition 4 of \cite{Manschot:2009ia},
invokes the standard Poisson resummation technique to establish the modular transformation property of the theta series.
Instead, our proof is based on the use of Vign\'eras' theorem \cite{Vigneras:1977}.
This theorem can be used to prove modularity for a quite general class of theta series,
and will play a crucial role in the study of Darboux coordinates relegated to \cite{abmp-to-appear}.

\subsection{Vign\'eras' theorem}

Let us start by stating Vign\'eras' theorem in a general fashion.
Let $\Lambda$ be an $n$-dimensional lattice equipped with a bilinear form
$B(\bfx,\bfy)\equiv \bfx\cdot\bfy$, where $\bfx,\bfy\in \Lambda \otimes \mathbb{R}$, such that its associated quadratic form
has signature $(n_+,n_-)$ and is integer valued, i.e. $\bfk^2\in\IZ$ for $\bfk\in\Lambda$.
Furthermore, let $\bfp\in\Lambda$  be a characteristic vector
(such that $\bfk^2+\bfk\cdot \bfp\in 2\mathbb{Z}$, $\forall \,\bfk \in \Lambda$),
$\bfmu\in\Lambda^*/\Lambda$ a glue vector, and $\lambda$ an arbitrary integer.
With the usual notation $q=\expe{\tau}$, we consider the following family of theta series
\be
\label{Vignerasth}
\vartheta_{\bfp,\bfmu}(\Phi,\lambda;\tau, \bfb, \bfc)=\tau_2^{-\lambda/2} \sum_{{\bfk}\in \Lambda+\bfmu+\hf\bfp}
(-1)^{\bfk\cdot\bfp}\,\Phi(\sqrt{2\tau_2}(\bfk+\bfb))\, q^{-\frac12 (\bfk+\bfb)^2}\,\expe{\bfc\cdot (\bfk+\haf\bfb)}
\ee
defined by the kernel $\Phi(\bfx)$.
Irrespective of the choice of this kernel and the parameter $\lambda$, any such theta series satisfies the
following elliptic properties
\be
\begin{split}
\label{Vigell}
\vartheta_{\bfp,\bfmu}\left({\Phi} ,\lambda; \tau, \bfb+\bfk,\bfc\right) =&(-1)^{\bfk\cdot\bfp}\,
\expe{-\haf\, \bfc\cdot \bfk} \vartheta_{\bfp,\bfmu}\left({\Phi} ,\lambda; \tau, \bfb,\bfc\right),
\\
\vphantom{A^A \over A_A}
\vartheta_{\bfp,\bfmu}\left({\Phi}, \lambda; \tau, \bfb,\bfc+\bfk \right)=&(-1)^{\bfk\cdot\bfp}\,
\expe{\haf\, \bfb\cdot \bfk} \vartheta_{\bfp,\bfmu}\left({\Phi} ,\lambda; \tau, \bfb,\bfc\right).
\end{split}
\ee

Now let us require that in addition the kernel satisfies the following two conditions:
\begin{enumerate}
\item
Let $D(\bfx)$ be any differential operator of order $\leq 2$, and
$R(\bfx)$ any polynomial of degree $\leq 2$. Then $f(\bfx)=\Phi(\bfx)\,e^{\pi\bfx^2/2}$ must
be such that $f(\bfx)$, $D(\bfx)f(\bfx)$ and $R(\bfx)f(\bfx)\in
L^2(\Lambda\otimes\mathbb{R})\bigcap L^1(\Lambda\otimes\mathbb{R})$.
\item
$\Phi(\bfx)$ must satisfy
\be
\label{Vigdif}
\left[ \partial_{\bfx}^2   + 2\pi  \bfx\cdot \pa_{\bfx}  \right] \Phi(\bfx)  = 2\pi \lambda\,  \Phi(\bfx) .
\ee
\end{enumerate}
Then the theta series \eqref{Vignerasth} transforms as a modular form of weight $(\lambda+n/2,0)$.
Explicitly the modular transformation properties are given by
\bea
\vartheta_{\bfp,\bfmu}\left( \Phi ,\lambda; -1/\tau, \bfc,-\bfb\right)
&=&\frac{(-\I\tau)^{\lambda+\frac{n}{2}}}{\sqrt{|\Lambda^*/\Lambda|}}\,
\expe{\tfrac14\, \bfp^2} \sum_{\bfnu\in\Lambda^*/\Lambda}
\expe{\bfmu\cdot \bfnu}
\vartheta_{\bfp,\bfmu}\left(\Phi ,\lambda; \tau, \bfb,\bfc\right),
\nn\\
\vartheta_{\bfp,\bfmu}\left( \Phi ,\lambda; \tau+1, \bfb,\bfc+\bfb)\right)
&=&\expe{-\tfrac12(\bfmu+\tfrac12 \bfp)^2}
\vartheta_{\bfp,\bfmu}\left( \Phi ,\lambda; \tau, \bfb,\bfc\right).
\label{eq:thetatransforms}
\eea
For a positive definite lattice (with $n_-=0$), the theta series
$\vartheta_{\bfp,\bfmu}$ is related to the standard one with
complex elliptic variables $\bfv\in \mathbb{C}^{n_+}$
under the change of variables
\be
\vartheta_{\bfp,\bfmu}\left(\tau, \bfb,\bfc\right) = e^{\I\pi (\tau\bfb^2-\bfb\cdot\bfc)}
\tilde {\vartheta}_{\bfp,\bfmu}\left(\tau, \bfv= \bfb \tau-\bfc\right).
\ee

One can check that the covariant derivatives preserve the form of the theta series \eqref{Vignerasth}
changing the kernel and parameter $\lambda$ according to
\be
\label{covdertheta}
\begin{split}
\tau_2^2 \partial_{\bar\tau}\ :\
& (\Phi,\lambda) \mapsto \(\frac{\I}{4} \( \bfx \partial_{\bfx} \Phi - \lambda \Phi\) , \lambda-2 \),
\\
\partial_{\tau}-\frac{\I(\lambda+\tfrac{n}{2})}{2\tau_2}\ :\ & (\Phi,\lambda)
\mapsto \(-\frac{\I}{4} \( \bfx \partial_{\bfx} \Phi + (\lambda+n+2\pi \bfx^2) \Phi\) , \lambda+2 \) .
\end{split}
\ee

\subsection{Examples: Siegel-Narain and Zwegers' theta series}

Let us now restrict to the case of signature $(1,n-1)$.
A useful class of solutions of \eqref{Vigdif} are functions of one variable, the projection
of $\bfx$ on a fixed time-like vector $\bft$ with $\bft^2>0$,
\be
\Phi(\bfx) = f(x_+^{(\bft)}) ,
\qquad
x_+^{(\bft)}=\frac{\bfx \cdot \bft}{\sqrt{\bft^2}} ,
\qquad
f''+2\pi (x_+^{(\bft)} f' - \lambda f) = 0 .
\label{eqf}
\ee
The solution $f=e^{-\pi (x_+^{(\bft)})^2}$ with $\lambda=-1$ gives, up to a factor of $\tau_2^{1/2}$,
the standard Siegel-Narain theta series \eqref{defth}, a modular form of weight $(\tfrac{n-1}{2},\tfrac12)$.

Another solution of \eqref{Vigdif} with $\lambda=0$
is provided by the function $f=E(x_+^{(\bft)})$ with $E(x)={\rm Erf}(\sqrt{\pi} x)$.
However, since $E(x)\to \sgn(x)$ as $x\to \infty$, it does not satisfy the decay conditions.
For two time-like vectors $\bft, \bft'$, however, the difference $f=E(x_+^{(\bft)})-E(x_+^{(\bft')})$ does.
This leads to Zwegers indefinite theta series \cite{Zwegers-thesis} of weight $(\tfrac{n}{2},0)$,
\bea
\widehat\Theta_{\bfp,\bfmu}(\tau,\bft,\bft',\bfb,\bfc)&=& \sum_{\bfk \in \Lambda+\bfmu+\tfrac12 \bfp}
\signkp
\left[ \Erf\left(\sqrt{2\pi \tau_2} \kbp^{(\bft)} \right)-
\Erf\left(\sqrt{2\pi \tau_2} \kbp^{(\bft')}  \right) \right]
\nn\\
&&\times
q^{-\tfrac12(\bfk+\bfb)^2}\,\expe{\bfc\cdot (\bfk+\haf\bfb)} ,
\label{Zwegthetahat}
\eea
which provides the modular completion of the holomorphic indefinite theta series
\bea
\Theta_{\bfp,\bfmu}(\tau,\bft,\bft',\bfb,\bfc)&=&\sum_{\bfk \in \Lambda+\bfmu+\tfrac12 \bfp}
\signkp\[\sgn( \kbp^{(\bft)} )-\sgn( \kbp^{(\bft')} )\]
\nn\\
&&\times
q^{-\tfrac12(\bfk+\bfb)^2}\,\expe{ \bfc\cdot (\bfk+\tfrac12\bfb)}.
\label{eq:indefTheta}
\eea

\subsection{Construction of $\widehat\Psi$}

Let us now consider a variant of \eqref{eq:indefTheta}, with an extra insertion of $\kbp^{(\bft)}$ in the
sum:
\be
\begin{split}
\label{eq:indefTheta2}
\Theta'_{\bfp,\bfmu}(\tau,\bft,\bft',\bfb,\bfc)=& \sum_{\bfk \in \Lambda+\bfmu+\tfrac12 \bfp}
 \signkp\[\sgn( \kbp^{(\bft)} )-\sgn( \kbp^{(\bft')} )\] \kbp^{(\bft)}\,
 \\ &
 \quad \times q^{-\tfrac12(\bfk+\bfb)^2}\,\expe{\bfc\cdot (\bfk+\tfrac12\bfb)}.
\end{split}
\ee
To find the modular completion of \eqref{eq:indefTheta2}, we need to find a solution of \eqref{Vigdif}
which asymptotes to
\be
\label{xpsgn}
x_+^{(\bft)} \[\sgn( x_+^{(\bft)} )-\sgn( x_+^{(\bft')} )\] .
\ee
The first term $|x_+^{(\bft)}|$ can be promoted to $F(x_+^{(\bft)})$ where
\be
\label{defF}
F(x)=x \, {\rm Erf}(\sqrt{\pi} x) + \frac{1}{\pi}\, e^{-\pi x^2},
\ee
which is a solution of \eqref{eqf} with $\lambda=1$. To deal with the second term, we decompose
$\bft$ into its projection on $\bft'$ and its orthogonal complement:
\be
\bft = \frac{\bft\cdot \bft'}{\bft'\cdot\bft'}\, \bft'+ \left[ \bft - \frac{\bft\cdot \bft'}{\bft'\cdot\bft'}\, \bft' \right].
\ee
Contracting with $\bfx/\sqrt{\bft^2}$ and multiplying by $\sgn( x_+^{(\bft')})$, one obtains
\be
x_+^{(\bft)} \sgn( x_+^{(\bft')}) = \frac{ \bft\cdot \bft'}{\sqrt{\bft^2\bft'^2}}\,
|x_+^{(\bft')}|
+  \left[ x_+^{(\bft)} -  \frac{ \bft\cdot \bft'}{\sqrt{\bft^2\bft'^2}}\, x_+^{(\bft')}
 \right]\sgn( x_+^{(\bft')}).
\ee
The first term can be promoted to $F(x_+^{(\bft')})$, while in the second term,
$\sgn( x_+^{(\bft')})$ can be promoted to $E(x_+^{(\bft')})$.
Thus, a solution of the Vign\'eras' equation with the required decay properties
can be obtained by promoting \eqref{xpsgn} to
\be
\begin{split}
\Phi(\bfx)= &F(x_+^{(\bft)}) -  \frac{ \bft\cdot \bft'}{\sqrt{\bft^2\bft'^2}}\,  F(x_+^{(\bft')})
- \left[ x_+^{(\bft)} -  \frac{ \bft\cdot \bft'}{\sqrt{\bft^2\bft'^2}}\, x_+^{(\bft')} \right] E(x_+^{(\bft')})
\\
= & x_+^{(\bft)} \left[ E(x_+^{(\bft)}) -  E(x_+^{(\bft')}) \right] +
\frac{1}{\pi} \,e^{-\pi (x_+^{(\bft)})^2} - \frac{ \bft\cdot \bft'}{\pi \sqrt{\bft^2\bft'^2}}\,
e^{-\pi (x_+^{(\bft')})^2}.
\end{split}
\label{complet-Phi}
\ee
By construction, this is a solution of \eqref{Vigdif} with $\lambda=1$ and thus, using this kernel in \eqref{Vignerasth},
one obtains a modular completion of \eqref{eq:indefTheta2} with weight $(\tfrac{n}{2}+1,0)$.
Note that unlike the case of Zwegers' theta series, the difference between \eqref{complet-Phi} and \eqref{xpsgn}
is {\it not} the difference of a function of $\bft$ and a function of $\bft'$.

We now apply this construction to produce the mock Siegel-Narain theta
series constructed in \cite{Manschot:2009ia}.
We start with a lattice $\Lambda\oplus\Lambda$
which carries the quadratic form $\cK^2=\kappa_{abc} k_{1}^a k_{1}^b p_1^c + \kappa_{abc} k_{2}^a k_{2}^b p_2^c$
of signature $(2,2b_2-2)$,
where $\cK=(\bfk_{1},\bfk_{2})$ and $\bfp_1,\bfp_2$ are two vectors in $\Lambda$ with $p_1^3,p_2^3>0$.
Let $\bfp=\bfp_1+\bfp_2$ and $\bft \in \Lambda \otimes \IR$, such that $t^3$, $(p_1t^2)$, $(p_2 t^2)$, $(p_1 p_2 t)$
are all positive.
Then it is easy to check that the vectors
\be
\cT = \frac{(t^a,t^a)}{\sqrt{(p t^2)}}\, ,
\qquad
\cP = \frac{(p_2^a,- p_1^a)}{\sqrt{(pp_1p_2)}}\, ,
\qquad
\cP' = \frac{((p_2 t^2) t^a, -(p_1t^2) t^a )}{\sqrt{(p_1t^2)(p_2t^2)(pt^2)}}
\ee
satisfy
\be
\begin{split}
\cT^2= \cP^2 = \cP'^2 = 1 ,&
\qquad
\cT\cdot\cP= \cT\cdot\cP' = 0,
\qquad
\cP\cdot\cP' = \sqrt{ \frac{(p t^2) (p_1 p_2 t)^2}{(p_1t^2)(p_2 t^2)(p p_1p_2)}},
\\
\cK\cdot\cP =&\, \frac{\langle\gamma_1,\gamma_2\rangle}{\sqrt{(pp_1p_2)}},
\qquad
\cK\cdot\cP' = \cI_{\gamma_1\gamma_2}(\bft,\bfb=0),
\end{split}
\label{vectors-rel}
\ee
where in the last two relations, as usual, we took the charge vectors as $\gamma_i=(0,p_i^a,\kappa_{abc}k_i^bp_i^c,q_{i,0})$.

Now we want to use the kernel \eqref{complet-Phi} with the above quadratic form and vectors $\bft,\bft'$ identified with
$\cP,\cP'$, respectively. However, the two cases differ by the signature of the quadratic form: due to the additional positive
direction, the naive use of \eqref{complet-Phi} with the above data leads to a kernel which does not decay in the direction
described by the vector $\cT$. Fortunately, the situation can be cured by multiplying by an additional exponential factor
where the charge vector is projected on $\cT$, as in the Siegel-Narain theta series.
In this way we arrive to the following kernel
\be
\Phi(\cK)= e^{-\pi (\cK\cdot\cT)^2} \! \left[
(\cK\cdot\cP) \Bigl( E(\cK\cdot\cP) -  E(\cK\cdot\cP') \Bigr) +
\frac{e^{-\pi (\cK\cdot\cP)^2} }{\pi} - \frac{(\cP\cdot\cP')\,e^{-\pi (\cK\cdot\cP')^2}}{\pi}
\right]\!.
\ee
This is a solution of Vign\'eras' equation with $\lambda=0$, which follows
from the fact that the two factors separately satisfy this equation with $\lambda=-1$ and $\lambda=1$, respectively,
and that $\cT$ is orthogonal to both $\cP$ and $\cP'$.
Furthermore, $\Phi(\cK)$ satisfies the required decay condition and thus it generates
a modular form of weight $(b_2,0)$,
\be
\label{Janth}
\widehat\Theta_{\bfp_1,\bfp_2,\bfmu_1,\bfmu_2}=\sum_{{\bfk_i}\in \Lambda_i+\bfmu_i+\hf\bfp_i}
(-1)^{\bfk_1\cdot\bfp_1+\bfk_2\cdot\bfp_2}\,\Phi(\sqrt{2\tau_2}(\cK+\cB))\, q^{-\frac12 (\cK+\cB)^2}\,\expe{\cC\cdot (\cK+\haf\,\cB)},
\ee
where $\cB=(\bfb,\bfb)$ and $\cC=(\bfc,\bfc)$. Using \eqref{vectors-rel}, it is straightforward to check
that this double theta series reproduces $\widehat \Psi_{\bfp_1,\bfp_2,\bfmu_1,\bfmu_2}$  introduced in section \ref{subsec-modul},
up to an overall factor of $-(-1)^{(p_1^2p_2)}\sqrt{8\tau_2/(pp_1p_2)}$. This completes the alternative proof
that this function is a modular form.

\section{Computing $R_{p,\mu}$}
\label{ap-R}

In this appendix we derive the explicit expression for the non-holomorphic completion $R_{\bfp,\bfmu}$
of the holomorphic mock modular form $h_{\bfp,\bfmu}$.
Our starting point is \eqref{hhtoRth}, where $\Psi^{(-)}_{\bfp_1,\bfp_2,\bfmu_1,\bfmu_2}$
is defined in \eqref{defPsih}, \eqref{ins-two}. First, we note the identity
\be
\label{IZid}
\cI_{\gamma_1\gamma_2}^2=(\bfq_1+\bfb)_{1+}^2+(\bfq_2+\bfb)_{2+}^2-(\bfq_1+\bfq_2+\bfb)_+^2
\ee
where the index 1 or 2 denotes the charge used to define the quadratic form, while the index $+$ denotes as usual
the projection on the K\"ahler modulus $\bft$. For instance,
$(\bfk)_{i+}^2=\frac{(ktp_i)^2}{(p_it^2)}$.
This identity can be used to rewrite the r.h.s. of \eqref{hhtoRth} as
\bea
&&-\frac{1}{4\pi}\sum_{\gamma_1,\gamma_2 \in \Gamma_+\atop \bfp_1+\bfp_2=\bfp}
(-1)^{\bfp_1\cdot\bfq_1+\bfp_2\cdot\bfq_2+(p_1^2p_2)}
\abs{\left<\gamma_1,\gamma_2\right>}\,
\bar \Omega_{\bfp_1,\bfmu_1}(\hat q_{0,1})
\bar \Omega_{\bfp_2,\bfmu_2}(\hat q_{0,2}) \,
\beta_{\frac{3}{2}}\!\left(\textstyle{\frac{2\tau_2}{(pp_1p_2)}}\, \langle\gamma_1,\gamma_2\rangle^2 \right)
\nn\\
& \times&
\expe{ - \tau (\hat q_{1,0}+
\hat q_{2,0})+\frac{\tau}{2}
\Bigl[ (\bfq_1+\bfq_2+\bfb)^2-(\bfq_1+\bfb)_1^2-(\bfq_2+\bfb)_2^2 \Bigr]}\cXt_{\bfp,\bfq_1+\bfq_2} .
\label{expr-for-R}
\eea

Next, we decompose each of the charges $\bfq_1$, $\bfq_2$ and $\bfq=\bfq_1+\bfq_2$
according to \eqref{defmu},
\be
\label{sumqis}
\begin{split}
q_{1,a}=&  \mu_{1,a}+ \frac12 \kappa_{abc} p_1^b p_1^c+\kappa_{abc} p_1^b \eps_1^c,
\\
q_{2,a} = & \mu_{2,a}+ \frac12 \kappa_{abc} p_2^b p_2^c+\kappa_{abc} p_2^b \eps_2^c,
\\
q_{a} =&  \mu_{a}+ \frac12 \kappa_{abc} p^b p^c+\kappa_{abc} p^b \eps^c ,
\end{split}
\ee
where $\bfeps_i,\bfeps \in\Lambda$ and $\tilde\bfmu_i \in\Lambda^\star/\Lambda_i$, $\tilde\bfmu\in\Lambda^\star/\Lambda$.
The  sum over charges in \eqref{expr-for-R} is then
\be
\sum_{\gamma_1,\gamma_2 \in \Gamma_+\atop \bfp_1+\bfp_2=\bfp}=
\sum_{\bfp_1+\bfp_2=\bfp}\sum_{\hat q_{i,0}}\sum_{\bfmu_i\in \Lambda^\star/\Lambda_i}\sum_{\bfeps_i\in\Lambda_i}\, .
\ee
Our goal is to exchange the sum over $\bfeps_1,\bfeps_2$ for a sum over $\bfeps,\bfmu$ and
the variable $\bfrho$ defined by
\be
\label{bor2}
\kappa_{abc} p_1^b \eps_1^c=\kappa_{abc} p_1^b \tilde\eps^c + \rho_a\, ,
\qquad
\kappa_{abc} p_2^b \eps_2^c=\kappa_{abc} p_2^b \tilde\eps^c - \rho_a\, .
\ee
Here $\tilde\bfeps$ is a non-integer vector which will be related to $\bfeps$ momentarily.
The equations \eqref{bor2} uniquely determine the pair $(\tilde\bfeps,\bfrho)$ for each pair $(\bfeps_1,\bfeps_2)$.
Next, we apply the standard decomposition to the sum of these two equations,
\be
\kappa_{abc} p_1^b \eps_1^c + \kappa_{abc} p_2^b \eps_2^c=
\kappa_{abc} p^b \tilde\eps^c  = \tilde\mu_a + \kappa_{abc} p^b \eps^c\, ,
\label{decomp-teps}
\ee
where $\tilde\bfmu\in\Lambda^\star/\Lambda$. Thus $\tilde\bfeps$ is
related to $\bfeps$ by
\be
\tilde\eps^a = \eps^a + \kappa^{ab}\tilde\mu_b \equiv \eps^a + \tilde\mu^a\, .
\label{teps}
\ee
Using \eqref{decomp-teps} in \eqref{sumqis}, we find that $\tilde\bfmu$
is related to $\bfmu$ by
\be
\tilde\mu_a =\mu_a-\mu_{1,a}-\mu_{2,a}+\kappa_{abc} p_1^b p_2^c .
\label{deftildemu}
\ee
As a result, we can now exchange the two vectors $\bfeps_i\in\Lambda_i$ for three variables: $\bfeps\in\Lambda$,
$\bfmu\in\Lambda^\star/\Lambda$ and $\bfrho$. As can be seen from \eqref{bor2} and \eqref{teps}, the latter is
such that
\be
\rho_a + \kappa_{abc} p_1^b \tilde\mu^c \in \Lambda_1,
\qquad
\rho_a-\kappa_{abc} p_2^b \tilde\mu^a \in \Lambda_2 .
\ee
Thus, one has
\be
\sum_{\bfeps_i\in\Lambda_i}=\sum_{\bfmu\in\Lambda^\star/\Lambda}\
\sum_{\bfk\in \Lambda+\bfmu+\hf\bfp}\
\sum_{\bfrho\in (\Lambda_1-\tilde\bfmu)\cap(\Lambda_2+\tilde\bfmu)}.
\label{tradesum}
\ee

Furthermore, let us substitute the decomposition of $\bfq_i$ in terms of the new variables into the combinations of charges
appearing in \eqref{expr-for-R}. There are three such combinations:
\begin{itemize}
\item
the symplectic product of two charges
\be
\Sgg \equiv \langle \gamma_1, \gamma_2\rangle =
p_2^a \mu_{1,a}-p_1^a \mu_{2,a}+\frac12\,\kappa_{abc} p_1^a p_2^b (p_2^c-p_1^c) + p^a \rho_a ;
\label{defSgg}
\ee
\item
the square bracket in the exponential
\be
(\bfq_1+\bfq_2+\bfb)^2-(\bfq_1+\bfb)_1^2-(\bfq_2+\bfb)_2^2=
{\rm Q}_{\bfp_1,\bfp_2}\left(\bfnu_1,\bfnu_2\right),
\ee
where the quadratic form ${\rm Q}_{\bfp_1,\bfp_2}:\Lambda^*\oplus \Lambda^*\to
\mathbb{Q}$ is defined by
\be
{\rm Q}_{\bfp_1,\bfp_2}(\bfq_1,\bfq_2)=(\bfq_1+\bfq_2)^2-(\bfq_1)_1^2-(\bfq_2)_2^2
\label{defQR}
\ee
and
\be
\begin{split}
\nu_{1,a}=&\, \mu_{1,a}+\hf\,\kappa_{abc}p_1^b(p_1^c+\tilde\mu^c)+\rho_a,
\\
\nu_{2,a}=&\, \mu_{2,a}+\hf\,\kappa_{abc}p_2^b(p_2^c+\tilde\mu^c)-\rho_a;
\end{split}
\label{defnui}
\ee
\item
the sign factor
\be
(-1)^{\bfp_1\cdot\bfq_1+\bfp_2\cdot\bfq_2+(p_1^2p_2)}=(-1)^{\bfk\cdot\bfp}(-1)^{\Sgg}.
\ee
\end{itemize}
Note that the only dependence on $\bfk$ appears in the sign factor and in $\cXt_{\bfp,\bfk}$.
Thus, the corresponding sum produces the theta series \eqref{defth} and one obtains
\be
\sum_{\bfmu\in\Lambda^\star/\Lambda} R_{\bfp,\bfmu}(\tau) \,\theta_{\bfp,\bfmu}(\tau,\bft,\bfb,\bfc)
\ee
with $R_{\bfp,\bfmu}(\tau)$ given in \eqref{defRmu}.

It is important to check that the sum \eqref{defRmu} is convergent. To this aim, note that for large $x$,
$\beta_\frac{3}{2}(x)< e^{-\pi x}$. Thus, we need to show
that
\be
\label{convcond}
(\bfrho)_1^2+(\bfrho)_2^2-\frac{2 (\bfp\cdot \bfrho)^2}{(p p_1 p_2)}<0.
\ee
Defining $\bfk_1,\bfk_2$ via $\rho_a=\kappa_{abc}p_1^bk^c_1=-\kappa_{abc}p_2^bk^c_2$,
\eqref{convcond} is equivalent to
\be
\label{convcond2}
\left[ (k_1^2 p_1)-\frac{(k_1p_1p)^2}{(pp_1p_2)}\right]+
\left[(k_2^2 p_2)-\frac{(k_2p_2p)^2}{(pp_1p_2)} \right] < 0.
\ee
Using $(pp_1p_2)<(p^2 p_1)$, one has
\be
(k_1^2 p_1)-\frac{(k_1p_1p)^2}{(p p_1p_2)}<(k_1^2 p_1)-\frac{(k_1p_1p)^2}{(p^2p_1)}\leq 0,
\ee
where the last inequality follows from $(k_1^2
p_1)-\frac{(k_1p_1p)^2}{(p^2p_1)}=(k_1)_-^2$ for $\bft=\bfp$. The
first bracket in \eqref{convcond2} is thus negative. Similarly,
the second bracket is negative. Thus, \eqref{convcond} holds, and the sum \eqref{defRmu}
is indeed absolutely convergent.

\section{Details on the contact potential}
\label{ap-cp}

\subsection{Calculation of $\delta e^\Phi$}
\label{ap-contact}

There are four sources of two-instanton terms in \eqref{phiinstmany}:
\begin{itemize}
\item
one-instanton contribution to $\cX_\gamma$ plugged in the integral term;
\item
one-instanton contribution to the mirror map for $u^a$ plugged in the central charge appearing in the same integral term;
\item
quadratic terms in the one-instanton contribution to the mirror map for $u^a$ coming from first `tree-level' term;
\item
two-instanton contribution to the mirror map for $u^a$ plugged in the first term.
\end{itemize}
Collecting all these contributions together and taking the large volume limit $t^a\to \infty$,
one arrives at the following result
\bea
\delta e^\Phi &=&-\frac{\tau_2}{16\pi^2}\sum_{\gamma \in\Gamma_+}\sigma_\gamma\bar\Omega(\gamma)
\int_{\ell_{\gammap}}\!\de z\!
\[q_0+q_a b^a+\frac{(bbp)}{2}
-2\I z(q_a t^a+(pbt))-\frac{3z^2(pt^2)}{2}\]\cXcl_{\gammap}\bigl(1+\cXq_{\gammap}\bigr)+{\rm c.c.}
\nn\\
&&
-\frac{1}{64\pi^4}\sum_{\gamma_1,\gamma_2\in\Gamma_+}
\sigma_{\gamma_1}\sigma_{\gamma_2}\bar\Omega(\gamma_1)\bar\Omega(\gamma_2)(tp_1p_2)
\(\Re\int_{\ell_{\gammap_1}}\!\!\de z_1\, \cXcl_{\gammap_1}\)\( \Re\int_{\ell_{\gammap_2}}\!\!\de z_2\, \cXcl_{\gammap_2}\).
\label{Phitwo-lv}
\eea
To further simplify this expression, we note the following identities
\bea
&&\ \ \sum_{\gamma\in\Gamma_+}\sigma_\gamma \bar\Omega(\gamma)\int_{\ell_{\gammap}} {\de z}
\(\frac{1}{4\pi\tau_2} -\I z\(q_a t^a+(pbt)\)-z^2(pt^2) \)\cXcl_{\gammap}
=0,
\\
&&\sum_{\gamma_1,\gamma_2\in\Gamma_+}\!\!\sigma_{\gamma_1}\sigma_{\gamma_2}\bar\Omega(\gamma_1) \bar\Omega(\gamma_2)
\int_{\ell_{\gammap_1}}\!\! {\de z_1}\int_{\ell_{\gammap_2}} \!\!{\de z_2}
\,\frac{\I\langle\gamma_1,\gamma_2\rangle}{z_2-z_1} \(\frac{1}{8\pi\tau_2}-\I z_1\(q_{1,a} t^a+(p_1 bt)\)-z_1^2(p_1t^2)\)
\cXcl_{\gammap_1}\cXcl_{\gammap_2}=0.
\nn
\label{totder}
\eea
The first one holds because the integrand appears to be a total derivative, while the second identity can be proven by symmetrizing
in charges and integrating by parts.
Then, substituting \eqref{cXone} into \eqref{Phitwo-lv} and using these identities,
the instanton contribution to the contact potential can be rewritten as
\bea
\delta e^\Phi &=&
-\frac{\tau_2}{16\pi^2}\,\sum_{\gamma\in\Gamma_+}\sigma_{\gamma}\bar\Omega(\gamma)
\int_{\ell_{\gammap}} {\de z}\,\cXcl_{\gammap}
\(\hat q_{0}+\frac12 (\bfq+\bfb)^2 -\frac{\I z}{2}\(q_{a} t^a+(pbt)\)-\frac{3}{8\pi\tau_2}\)
\nn\\
&&
+\frac{\tau_2}{32\pi^3}\,\sum_{\gamma_1,\gamma_2\in\Gamma_+}\sigma_{\gamma_1}\sigma_{\gamma_2}\bar\Omega(\gamma_1) \bar\Omega(\gamma_2)
\int_{\ell_{\gammap_1}} {\de z_1}\,\cXcl_{\gammap_1}\int_{\ell_{\gammap_2}} {\de z_2}\,\cXcl_{\gammap_2}\[
\frac{(tp_1p_2)}{16\pi\tau_2}
\right.
\nn\\
&& \left.\quad
-\((tp_1p_2)-\frac{\I\langle\gamma_1,\gamma_2\rangle}{z_2-z_1}\) \(\hat q_{1,0}+\hf\,(\bfq_1+\bfb)^2
-\frac{\I z_1}{2}\(q_{1,a} t^a+(p_1 bt)\)-\frac{3}{16\pi\tau_2}\)\]+{\rm c.c.}
\nn\\
&&
-\frac{1}{8}\sum_{\bfp_1,\bfp_2}(tp_1p_2)\,\cFq_{\bfp_1}\overline{\cFq_{\bfp_2}},
\label{Phitwo-lvff}
\eea
where the function $\cFq_{\bfp}$ is defined in \eqref{defcFq}.
Using
\be
\cD_{\wh}\cXcl_{\gamma} =-\(\hat q_0+\hf\,(\bfq+\bfb)^2
-\frac{\I z}{2}\(q_a t^a+(pbt)\)+\frac{\wh}{4\pi\tau_2}\)\cXcl_{\gamma},
\ee
it is straightforward to check that this result is equivalent to the representation \eqref{Phitwo-Dp}.

\subsection{Calculation of the double integral}
\label{ap-doubleintcont}

Here, we provide the details of computation of the double integral $\cY_{\gamma_1\gamma_2}$ defined in \eqref{defdInt}.
We write it in the following general form
\be
\label{defdInt-gen}
\dInt=\int_{\ell_{\gamma_1}}\!\! \de z_1 \int_{\ell_{\gamma_2}} \!\!\de z_2\,
\frac{e^{-2\pi\tau_2 \(a_1\(z_1+\frac{\I b_1}{a_1}\)^2+a_2\(z_2+\frac{\I b_2}{a_2}\)^2\)}}{z_2-z_1}\, ,
\ee
where the contours $\ell_{\gamma_i}$ in the $z$-plane are arcs running from $-1$ to $1$ and passing through $-\I b_i/a_i$.
We are interested in the limit $a_i\gg 1$, which corresponds to the large volume limit on $\cM_H$.
Then one can deform the contours into straight lines $\IR -{\I b_i}/{a_i}$
since this changes the integral by an exponentially small contribution and, importantly,
we do not pick up any residue while doing this.
Performing the change of integration variables
\be
z_1= -\frac{\I b_1}{a_1}+v-\frac{a_2 u}{a_1+a_2},
\qquad
z_2= -\frac{\I b_2}{a_2}+v+\frac{a_1 u}{a_1+a_2},
\ee
one finds that the integral becomes
\be
\label{dInt-calc}
\dInt=\int_{\IR}\de v\, e^{-2\pi\tau_2(a_1+a_2)v^2}
\int_{\IR}\de u\,\frac{e^{-2\pi\tau_2\,\frac{a_1a_2}{a_1+a_2}\,u^2}}{u+\I\(\frac{b_1}{a_1}-\frac{b_2}{a_2}\)}\, .
\ee
The first factor is Gaussian, whereas the second can be evaluated using the formula
\be
\label{ztov}
\int_{\IR} \frac{dx}{x-\I \alpha}e^{-\beta^2 x^2}
=
\I\pi\, \sgn({\rm Re}(\alpha))\,e^{\alpha^2\beta^2}\, {\rm Erfc}\!\left(\sgn({\rm Re}(\alpha\beta))\alpha \beta \right).
\ee
For \eqref{dInt-calc}, this gives
\be
\dInt=-\frac{\I\pi\, \sgn(a_2 b_1-a_1 b_2)}{\sqrt{2\tau_2(a_1+a_2)}}\, e^{2\pi\tau_2\,\frac{(a_2 b_1-a_1 b_2)^2}{a_1a_2(a_1+a_2)}}
\beta_{\half}\(2\tau_2\,\frac{(a_2 b_1-a_1 b_2)^2}{a_1a_2(a_1+a_2)}\).
\ee
Substituting now $a_i=(p_it^2)$ and $b_i=(q_{i,a}+(p_ib)_a)t^a$, and noting that
$\frac{a_2b_1-a_1b_2}{\sqrt{a_1a_2(a_1+a_2)}}=\cI_{\gamma_1\gamma_2}$, one
reproduces the result \eqref{defdIntcp}.


\providecommand{\href}[2]{#2}\begingroup\raggedright\endgroup

\end{document}